\documentclass[]{spie}  

\usepackage{amsmath,amsfonts,amssymb}
\usepackage{graphicx}
\usepackage[colorlinks=true, allcolors=blue]{hyperref}
\usepackage{siunitx}
\usepackage{comment}
\usepackage{placeins}
\usepackage{subcaption}

\title{WST Multi-Object Spectrograph Fiber Positioners: Development of a 32,000-Unit Precision Robotic System}

\author[a]{Sébastien Pernecker}
\author[a]{Maxime Rombach}
\author[a]{Malak Galal}
\author[a]{Jonathan Wei}
\author[a]{Oliver Pineda Suárez}
\author[b]{David Lee}
\author[b]{Steve Watson}
\author[b]{Younes Chahid}
\author[b]{Chris Waring}
\author[b,c]{Anmol Goyal}
\author[d]{Joseph W. Barrow}
\author[d]{Will Saunders}
\author[d]{Jon Lawrence}
\author[e]{Aaron Omadutt}
\author[e]{Roelof S. de Jong}
\author[a]{Jean-Paul Kneib}

\affil[a]{Institute of Physics, Laboratory of Astrophysics, Ecole Polytechnique Federale de Lausanne (EPFL), Observatoire de Sauverny, CH-1290 Versoix, Switzerland}
\affil[b]{STFC UK Astronomy Technology Centre, Edinburgh, United-Kingdom}
\affil[c]{STFC Rutherford Appleton Lab, Didcot, United Kingdom}
\affil[d]{Australian Astronomical Optics (AAO), Macquarie University, NSW 2109, Australia}
\affil[e]{Leibniz-Institut für Astrophysik Potsdam (AIP), An der Sternwarte 16, 14482 Potsdam, Germany }

\authorinfo{Further author information: (Send correspondence to S. P.)\\S. P.: E-mail: sebastien.pernecker@epfl.ch, Telephone: +41 21 693 97 24}

\begin{document} 
\maketitle

\begin{abstract}
The Wide-field Spectroscopic Telescope Multi-Object Spectrograph requires an unprecedented fiber positioning system comprising 30,000 low-resolution and 2,000 high-resolution positioners across a 3.1 deg² field of view. Each robotic positioner must achieve 5 $\mu$m RMS positioning accuracy in a densely packed focal plane, representing a more than sixfold scale increase over current instruments like 4MOST and DESI. To mitigate risks associated with industrial-scale production of 32,000 precision mechanisms, WST is pursuing a multi-concept development strategy. Four distinct positioner architectures are being prototyped and tested by an international collaboration (EPFL, AIP, UKATC, AAO) using 6.2 mm pitch triangular modules of 63 units each or a new inline modular concept. Performance metrics including positioning accuracy, repeatability, reconfiguration speed, collision avoidance, and manufacturability are being systematically evaluated. Down-selection to one or two concepts is planned for 2026-2027 during the HORIZON Europe-funded conceptual study phase. Current prototype testing demonstrates feasibility of meeting specifications, supporting WST's path toward first light in the early 2040s as ESO's next major spectroscopic facility.
\end{abstract}


\section{INTRODUCTION}

Multi-object spectroscopy (MOS) is among the most productive observational techniques in modern astronomy, enabling the simultaneous acquisition of spectra from hundreds to thousands of sources across a wide field of view. 

Answering the most pressing questions in contemporary astrophysics, from the nature of dark energy driving cosmic acceleration, to the interplay between gas, stars, and dark matter in galaxies across cosmic time, to the detailed chemical enrichment history of the Milky Way and its satellites, requires spectroscopic datasets of unprecedented statistical completeness over large sky areas \cite{Mainieri2024}. Deep imaging surveys from facilities such as Euclid, the Vera C. Rubin Observatory, and the Roman Space Telescope will identify billions of targets; however, realising their full scientific potential demands matching spectroscopic follow-up capacity that no existing facility can provide.

The replacement of manually plugged fibre plates with robotic fibre positioners has been transformative: it eliminates field-change dead time, allows rapid focal-plane reconfiguration between exposures, and makes statistically complete surveys of large cosmic volumes practically feasible. First-generation robotic MOS systems are now fully operational, DESI deploys 5{,}000 positioners across a 3.2~deg$^{2}$ field to conduct a five-year dark-energy survey \cite{DESI}, while 4MOST will field approximately 2{,}400 positioners for its multi-survey programme at VISTA~\cite{4MOST}. These instruments have established the technology readiness and operational playbook upon which the next generation will build.

The Wide-field Spectroscopic Telescope (WST) is ESO's proposed flagship spectroscopic facility, targeting first light in the early 2040s~\cite{WST}. Its science case demands an unprecedented focal-plane multiplexing capability: 30{,}000 low-resolution and 2{,}000 high-resolution fibre positioners tiling a 3.1~deg$^{2}$ field of view, for a total of 32{,}000 robotic units, more than a sixfold increase over current instruments. The 30,000 low-resolution fibres serve the primary cosmological and extragalactic survey programmes, where large target statistics at moderate spectral resolution $(r \sim 5000)$ are paramount. The 2000 high-resolution fibres $(R \sim 20000-40000)$ address Galactic archaeology and stellar physics goals requiring detailed chemical abundance measurements and precise radial velocities.

This multiplexing capacity is set by the WST science case \cite{Mainieri2024,Bacon2024} : mapping the cosmic web and galaxy evolution out to redshift $z \sim 5$ with statistically complete sampling, constraining neutrino masses and primordial non-Gaussianity through baryon acoustic oscillation and redshift-space distortion measurements, and resolving stellar populations in the Milky Way and Local Group each independently require target densities that only 20,000+ simultaneously deployed fibres can achieve within realistic survey timescales.

Each positioner must achieve 5~$\mu$m RMS positioning accuracy within a 6.2~mm centre-to-centre pitch, yielding a fibre density 2.8 times higher than the state-of-the-art. At this scale, the focal plane system alone spans a 1{,}500~mm diameter curved surface, requires approximately 520 independent modules, and must sustain reliable operation over a target lifetime of 20 years. Assembly, integration, metrology, maintenance strategy, and production cost all become first-order design drivers alongside raw positioning performance.

Producing 32{,}000 precision robotic mechanisms at industrial scale introduces engineering risks that no single design concept can be assumed to mitigate in advance. To manage this, WST has adopted a deliberate multi-concept development strategy within its Work Package~4.2, engaging an international consortium (EPFL, AIP, UKATC, AAO) to prototype and evaluate four distinct positioner architectures in parallel: a theta-phi SCARA robot, a FLEX nitinol-tube concept, an R-theta architecture, and a tilting-spine design. Two modular arrangements, triangular and inline curvilinear,  are likewise being assessed in parallel. Performance metrics including positioning accuracy, repeatability, patrol coverage, collision avoidance, reconfiguration speed, and manufacturability are being systematically characterised against a common set of requirements. A down-selection to one or two preferred concepts is planned for 2026-2027 during the HORIZON~Europe-funded conceptual study phase, with the selected designs then advanced toward preliminary design.

This paper reports the status of the WST focal-plane positioner development programme. Section~\ref{sec:positioners} describes each of the four positioner concepts, with emphasis on the theta-phi SCARA design developed at EPFL and the prototype testing results obtained to date. Section~\ref{sec:ModularityConcept} presents the two competing modular focal-plane architectures. Section~\ref{sec:TradeOffStudy} describes the structured trade-off methodology, combining the Analytic Hierarchy Process for criteria weighting with a Pugh matrix for comparative scoring, that will guide concept down-selection. Section~\ref{sec:Conclusion} summarises current status and outlines next steps toward the WST conceptual design review.

\section{POSITIONERS}
\label{sec:positioners}
\subsection{MULTIPLE DESIGNS APPROACH}
\subsection{FLEX Positioner}
\begin{figure}[ht]
    \centering
    \includegraphics[width=\linewidth]{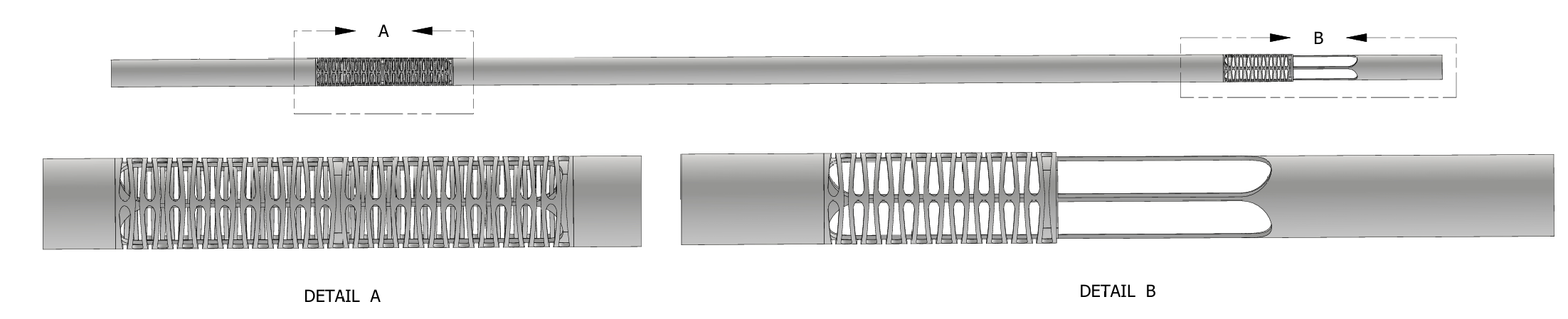}
    \caption{FLEX Positioner tubes with detail views of the lower lattice cutout (detail A) and fiber tip flexure (detail B) }
    \label{fig:FLEX_positioner}
\end{figure}

The Fiber Location EXtender (FLEX) is a novel fiber positioning concept currently being developed at the Leibniz-Institut für Astrophysik Potsdam (AIP). The positioner design consists of using three concentric nitinol tubes, each featuring laser-cut stress-reducing flexures as represented on Figure \ref{fig:FLEX_positioner}. 

\FloatBarrier

This nested configuration and geometric flexures allows the positioner tip to translate across its patrol area while maintaining high degrees of telecentricity relative to the focal plane. Due to the superelastic material properties of nitinol, the FLEX system is able to withstand large deflections without incurring any permanent deformation allowing the system to maintain an extremely high cycle life.

Additionally, the hollow tube construction provides a protected volume for fiber routing, allowing a 200 micron fiber to run freely within the body, reducing fiber curvature during actuation thus preventing localized mechanical stress and minimizing potential focal-ratio degradation effects (FRD).

\begin{figure}[ht]
    \centering
    \includegraphics[width=\linewidth]{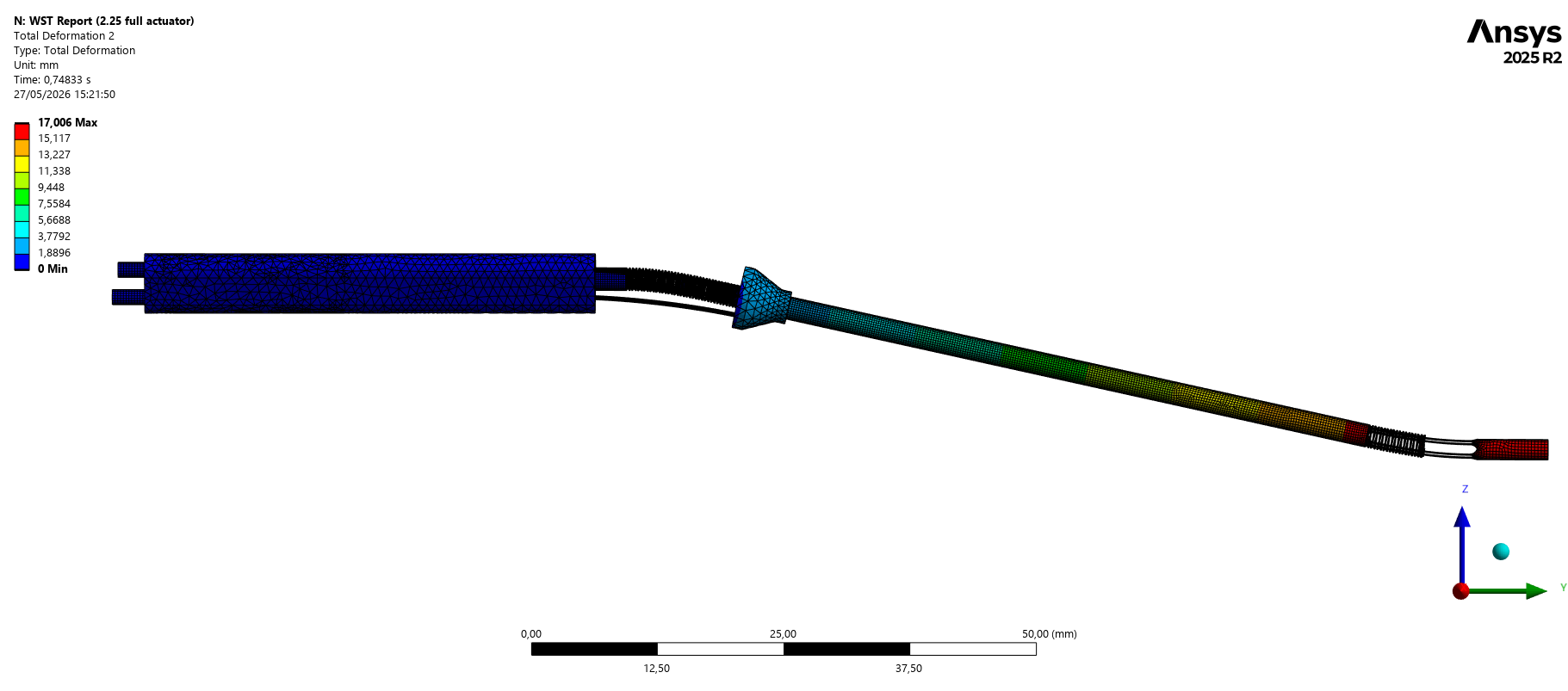}
    \caption{Flex Positioner deformation simulation}
    \label{fig:FLEX_pos_ANSYS}
\end{figure}

\begin{figure}[ht]
    \centering
    \begin{subfigure}[b]{0.48\textwidth}
        \centering
        \includegraphics[width=\textwidth]{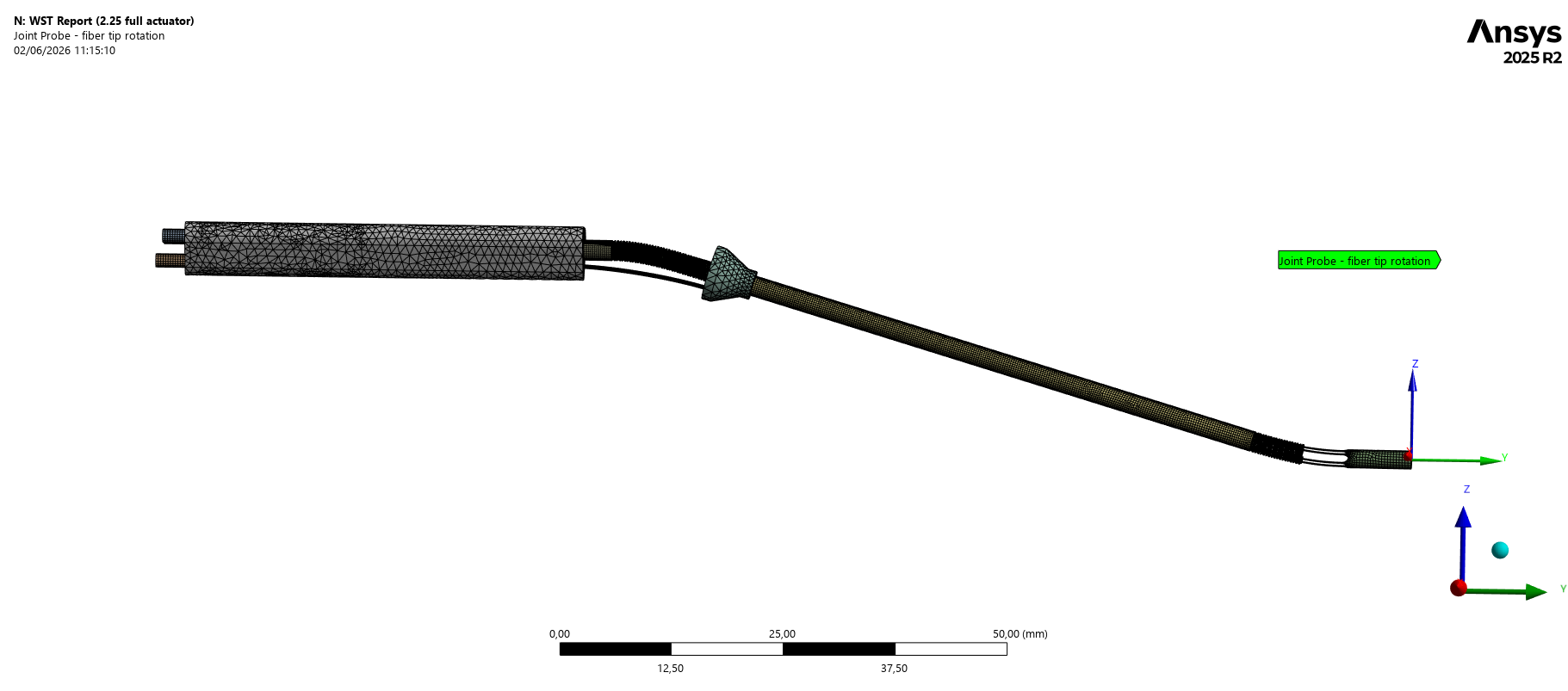}
        \caption{FLEX FEA model displaying fiber tip rotation at defined coordinate selection}
        \label{fig:FLEX - fiber tip rotation}
    \end{subfigure}
    \hfill 
    \begin{subfigure}[b]{0.48\textwidth}
        \centering
        \includegraphics[width=\textwidth]{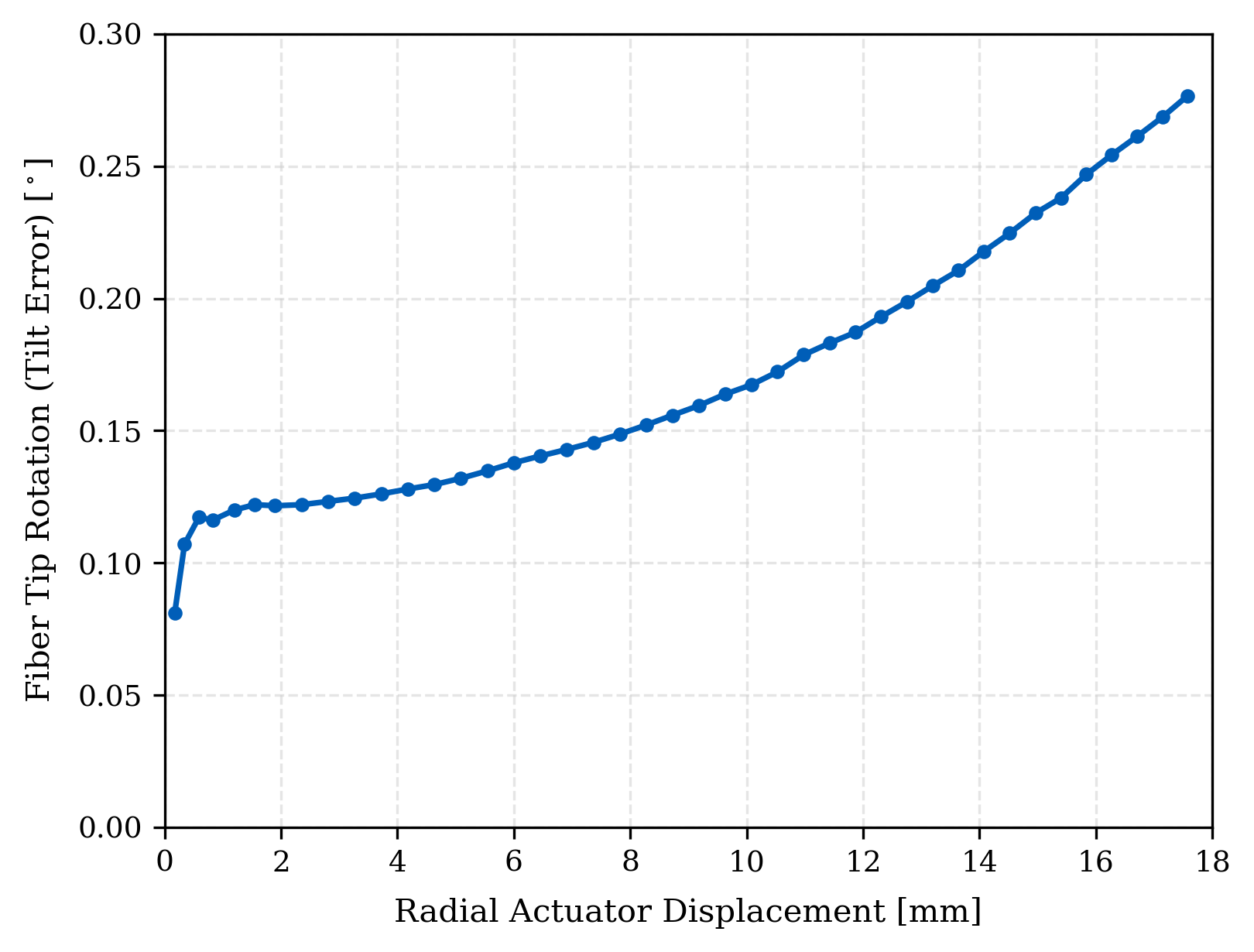}
        \caption{Plot of fiber tilt error vs. displacement}
        \label{fig:rotation vs displacement plot}
    \end{subfigure}
    \vspace{10pt}
    \caption{Evaluation of the FLEX positioner telecentricity error over its patrol radius. The localized deformation at the fiber tip probe (a) produces a non-linear tilt error (b) that reaches a maximum of $\approx 0.277^{\circ}$ at $\approx$17.58~mm} displacement
    \label{fig:combined_telecentricity}
\end{figure}

Driven by three piezo motors in a tripod configuration, the FLEX system provides high-resolution XY translation as well as active Z-axis refocusing. This setup allows the fiber tip to remain in-line with the focal plane throughout its entire patrol range. With a body length of 110mm, the FLEX positioner, when fully actuated, achieves a patrol radii of ~17mm (Figure \ref{fig:FLEX_pos_ANSYS}) with a telecentric error of $<$0.3 degrees as represented on Figure \ref{fig:combined_telecentricity}. 

\FloatBarrier

\FloatBarrier

The large patrol radii ($>$2.5x pitch) allows the FLEX positioner, when paired with a High Resolution (HR) fiber configured in a 1:16 packing structure, to achieve full focal surface coverage. Furthermore, the inherent self-locking characteristic of piezo actuators allows the fiber tip to hold its position without power after translating, improving power efficiency during long-exposure observations. Designed for the high multiplex MOS system of WST, the positioner can be paired with both in-line and triangular modules being developed, occupying very compact footprints, allowing for highly efficient tiling in focal surface assemblies.
\subsection{R-theta Positioner concept}
\label{sec:ukatc concept intro}
The UKATC WST positioner concept uses an R-theta architecture\cite{louth2024flexure}, in which each positioner rotates about a central axis and then extends radially to reach its target position. The centre-to-centre spacing between positioners (pitch) is 7.5~mm, and each positioner has an individual reach (patrol diameter) of more than 15~mm, providing full field coverage with a minimum of two fibres per location and built-in redundancy against positioner failures. In the current flexure design, FEA simulation at 7.5~mm radial extension show a defocus value of less than $50\,\mu\mathrm{m}$ (Figure ~\ref{fig:ukatc-FEA}) and pupil telecentricity of less than 0.4 degrees. The total estimated optical transmissive loss of this concept in operation is 6.78\%.

A radial architecture has an intrinsic advantage over current state-of-the-art theta-phi positioning systems in terms of path planning simplicity. However, the UKATC concept seeks to leverage this benefit further by adding on-board position sensing. This enables closed-loop positioning, making each positioner operationally independent of the focal plane metrology system and spectrographs (not requiring fibre backlighting). 
This positioner architecture could therefore have a major impact on the operational functionality of WST, improving the flexibility and efficiency of focal plane reconfiguration. It would allow a quick and robust path planning system which is tolerant to fault conditions and can accommodate last-minute Targets of Opportunity (ToO). It is also possible to theorise novel functionality, like reconfiguring high-resolution targets while simultaneously observing in low-resolution, or adjusting fibre positions to track moving targets in the field.

\begin{figure}[ht]
    \centering
    \includegraphics[width=\linewidth]{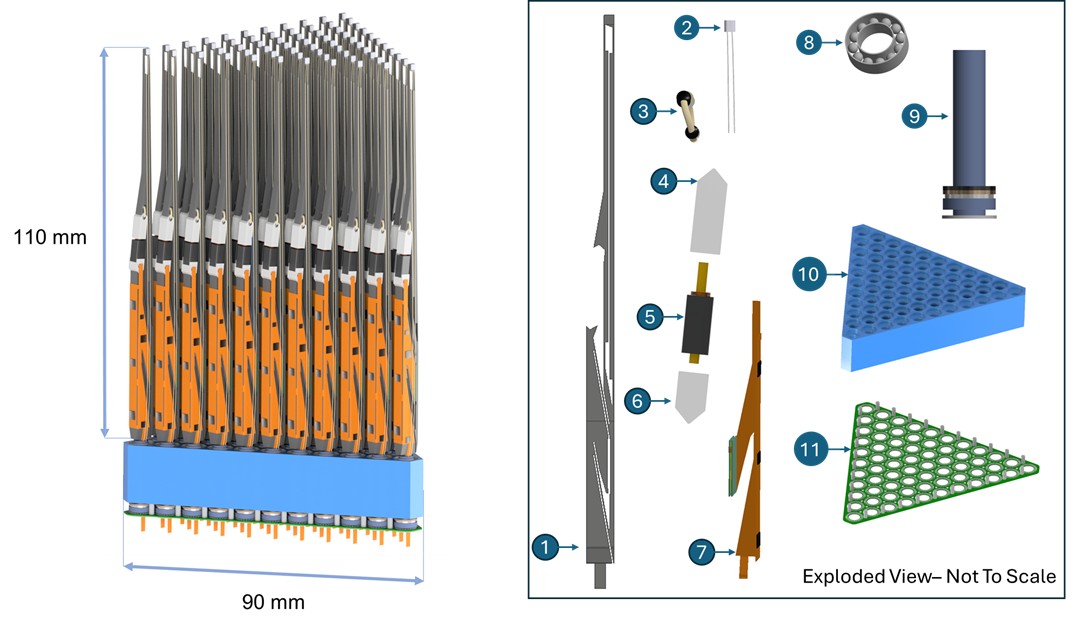}
    \caption{UKATC R-theta positioner concept; 
1. Radial flexure, 2. Fibre button, 3. Actuator ligament, 4. Top nut, 5. Radial actuator, 6. Bottom nut, 7. Flexi-rigid interface PCB, 8. Theta bearing, 9. Spindle assembly, 10. Baseplate, 11. Theta motor PCB
(right)}
    \label{fig:UKATC concept render}
\end{figure}\FloatBarrier

\subsubsection{Flexure design}
\begin{figure}[ht]
    \centering
    \includegraphics[width=.75\linewidth]{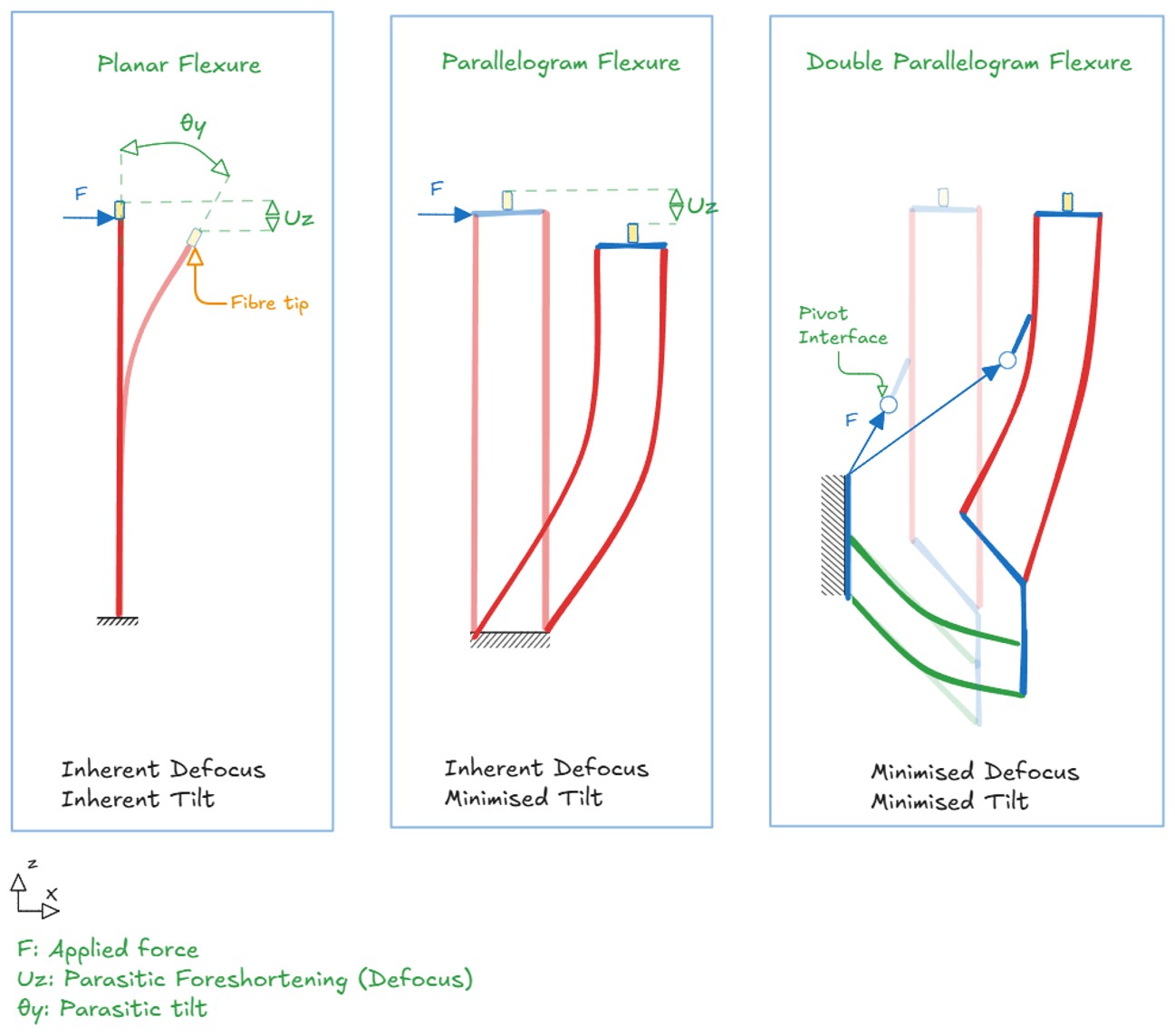}
    \caption{Impact of flexure geometry on defocus and tilt errors.}
        \label{fig:ukatc-drawing}
\end{figure}

\begin{figure}[ht]
    \centering
    \includegraphics[width=\linewidth]{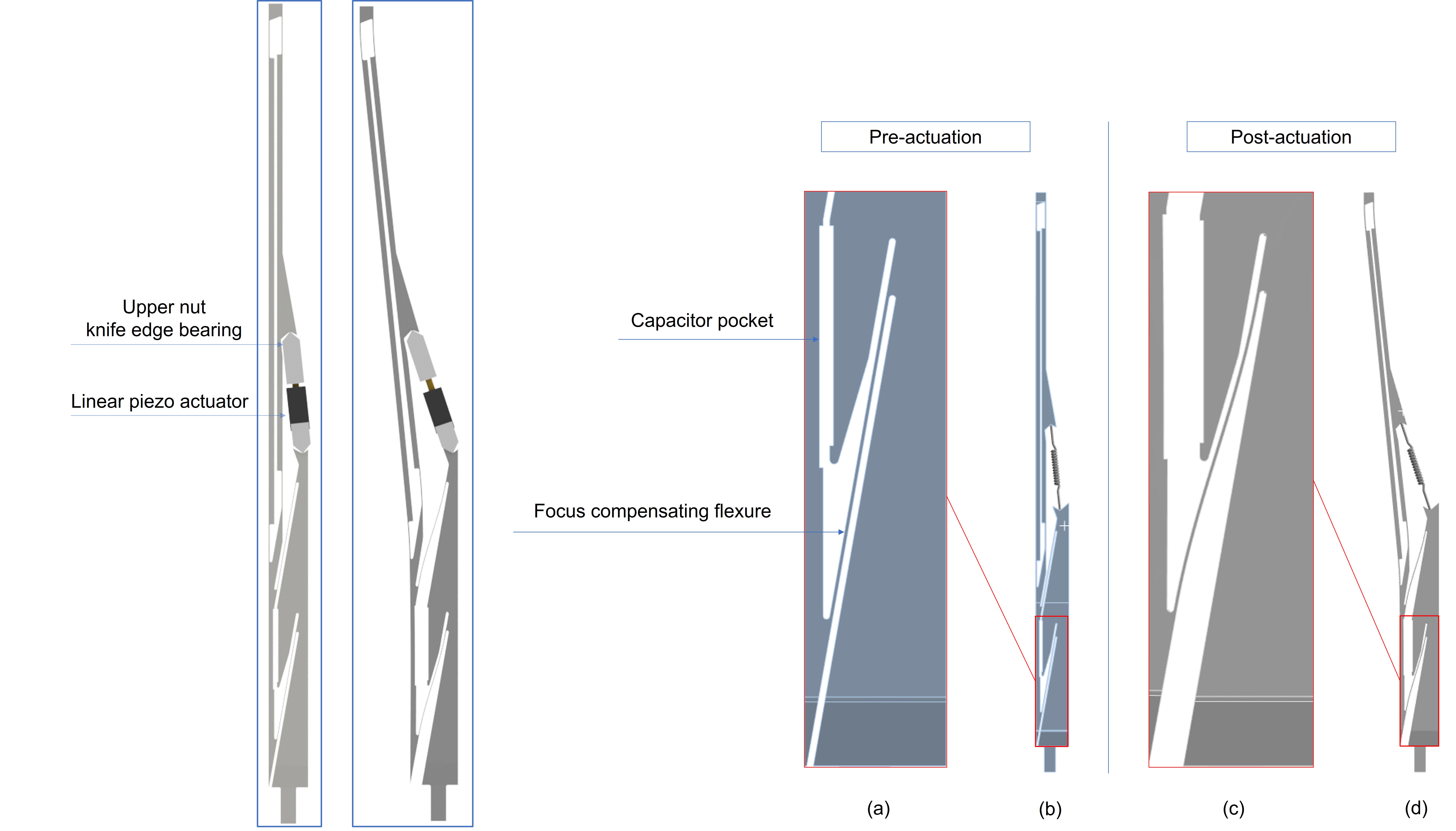}
    \caption{Renders of the positioner in the resting and actuated positions, with compensator details (right)}
        \label{fig:ukatc prepost-actuation}
\end{figure}

\begin{figure}[ht]
    \centering
    \includegraphics[width=\linewidth]{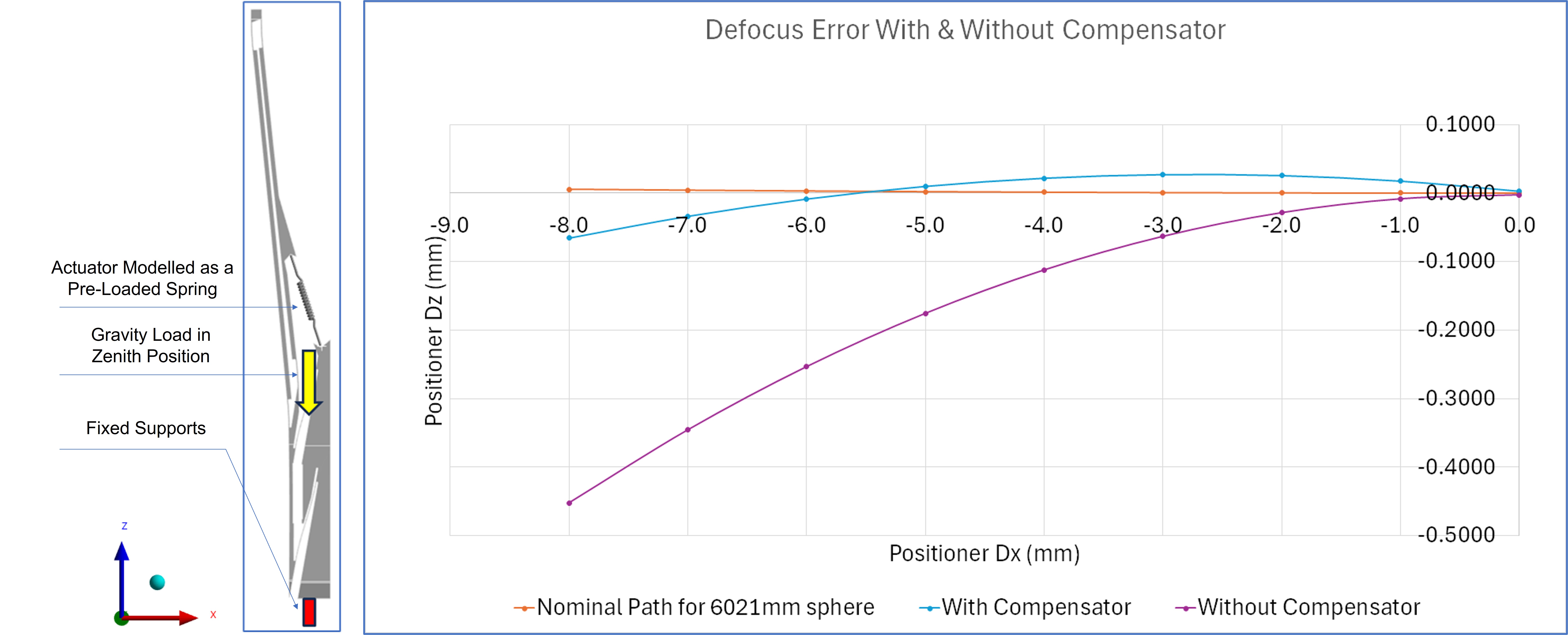}
    \caption{XY motion of single parallelogram flexure (without compensator) and double parallelogram flexure (with compensator).}
        \label{fig:ukatc-FEA}
\end{figure}

The UKATC positioner’s radial motion uses a focus-compensated double-parallelogram flexure. A single parallelogram does not produce perfectly straight motion and exhibits foreshortening, which causes fibre defocus. To address this, an angled compensating parallelogram is added at the base of the flexure. Under actuation, it introduces a positive piston component that partly offsets foreshortening and helps keep defocus within acceptable limits, as seen in Figure ~\ref{fig:ukatc-drawing}. The compensator flexure also provides a convenient location for measuring deflection, which can then be used to determine the fibre tip radial position. The proposed approach is to integrate a capacitance sensor that measures the separation between two electrodes within the compensator geometry, named as the the capacitor pocket location in Figure ~\ref{fig:ukatc prepost-actuation}.

\FloatBarrier

\FloatBarrier

\FloatBarrier

\subsubsection{Manufacturing and assembly}
The flexure is proposed to be manufactured from 7055 T77511 aluminium alloy due to an exceptional ratio of yield strength to modulus and fatigue strength. The flexure geometry is designed to be effectively 2-dimensional, allowing the adoption of micro-waterjet cutting as the primary manufacturing process. This process has part tolerances ($\pm 0.01$~mm) and minimum part thicknesses (0.1~mm) similar results to the wire-EDM process but is quicker and more economic for large batch quantities and is therefore a prudent choice for WST. 
The fibre is proposed to be mounted in a ceramic button which glues to the side wall of the flexure, with the fibre running freely down the length of the flexure and feeding through the spindle. As such, the flexure is manufactured with an asymmetric wedge profile on one side, meaning that the fibre sits at the central axis. This geometry is produced prior to the flexure profile cutting and does not interfere with the water-jet process.

\subsubsection{Radial actuation}
\begin{figure}[ht]
    \centering
    \includegraphics[width=0.75\linewidth]{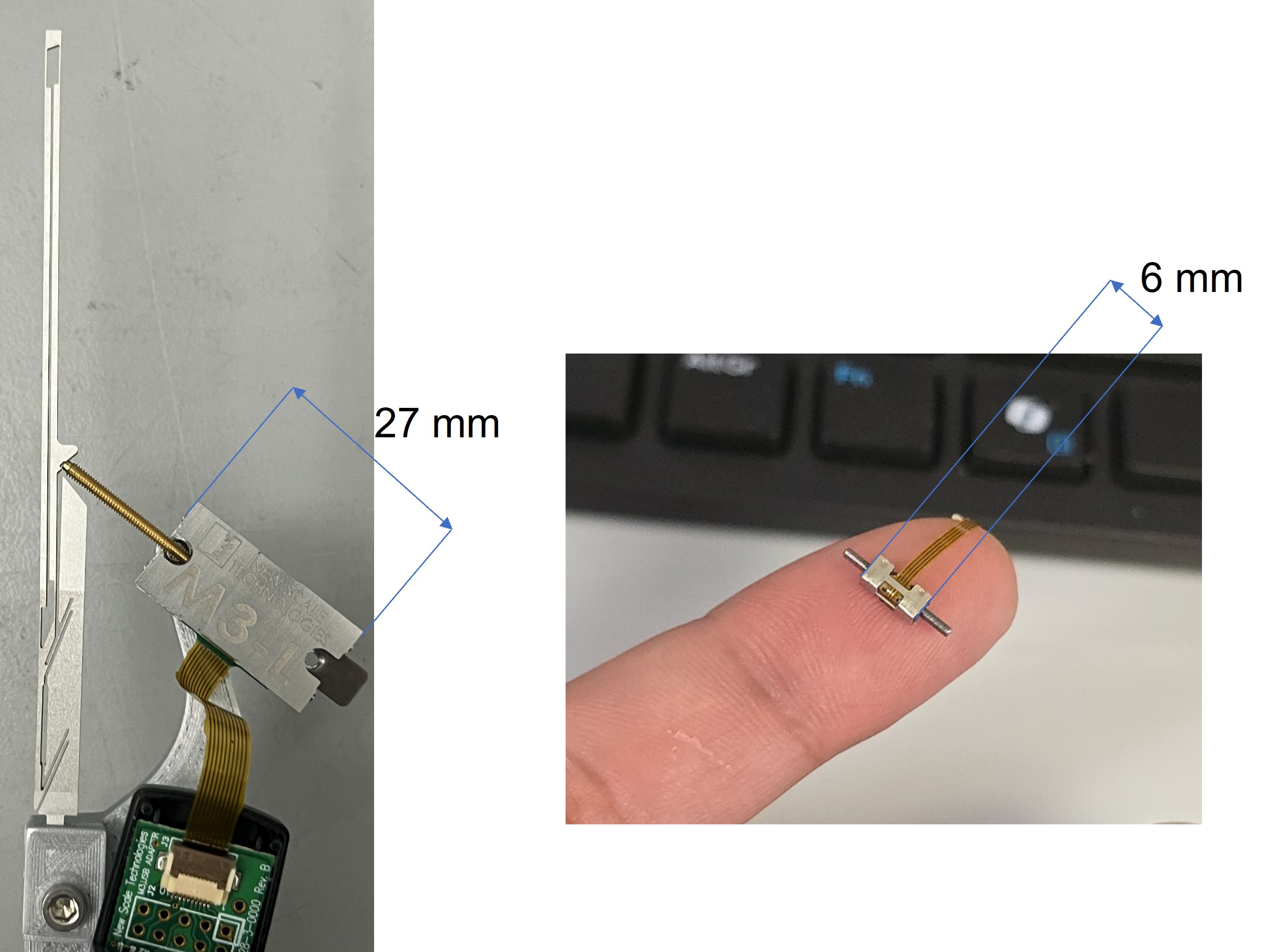}
    \caption{Flexure prototype made from 6082 T6 aluminium via EDM wire cut process assembled with a Newscale M3-L encompassing a SQL-RV-1.8 and closed-loop sensing (left) ; Newscale SQL-RV-1.8 linear piezo actuator (right).}
        \label{fig:ukatcradialactuation}
\end{figure}

The radial flexure is actuated by a piezo-driven lead screw, currently prototyped with a NewScale SQL-RV-1.8 squiggle motor seen in Figure ~\ref{fig:ukatcradialactuation} (right). The current design amplifies actuator motion 8× at the fibre, so achieving $10\,\mu\mathrm{m}$ accuracy at the fibre tip requires $0.5\,\mu\mathrm{m}$ actuator resolution. The Squiggle motor provides submicron resolution, with overall accuracy limited by the proposed capacitive position sensor. It is also compact, offers sufficient range and speed, and holds position without input power. The motor is mounted between two knife edge bearing nuts which allow rotation as the flexure displaces; the lower nut fixes to the body of the motor and the top one provides a captive interface into which the rotating lead screw can be converted into a linear force. The whole assembly is compressed within the flexure – this requires that the flexure is manufactured at a ‘negative’ position. The compression force is important to the functioning of the piezo motor, as the lead screw must be preloaded into the driving piezo ‘nut’.

\FloatBarrier

\subsubsection{Theta actuation}

The radial flexure would be mounted to a hollow spindle which rotates in two bearings mounted in the baseplate. The spindle would be actuated by a piezo ultrasonic ring motor. This motor uses a ring of piezo crystals which produce a standing wave to rotates a pre-loaded rotor. They have exceptional resolution, hold position with no power, no backlash and are very compact.
No commercially available motors meet the specific WST requirements but discussions are underway with several suppliers about producing custom motor units or, ideally, an array of motors on a common PCB, as seen in Figure ~\ref{fig:UKATC concept render} (right, 11), with incorporated capacitive sensors.

\subsubsection{Prototyping}

\begin{figure}[ht]
    \centering
    \includegraphics[width=1\linewidth]{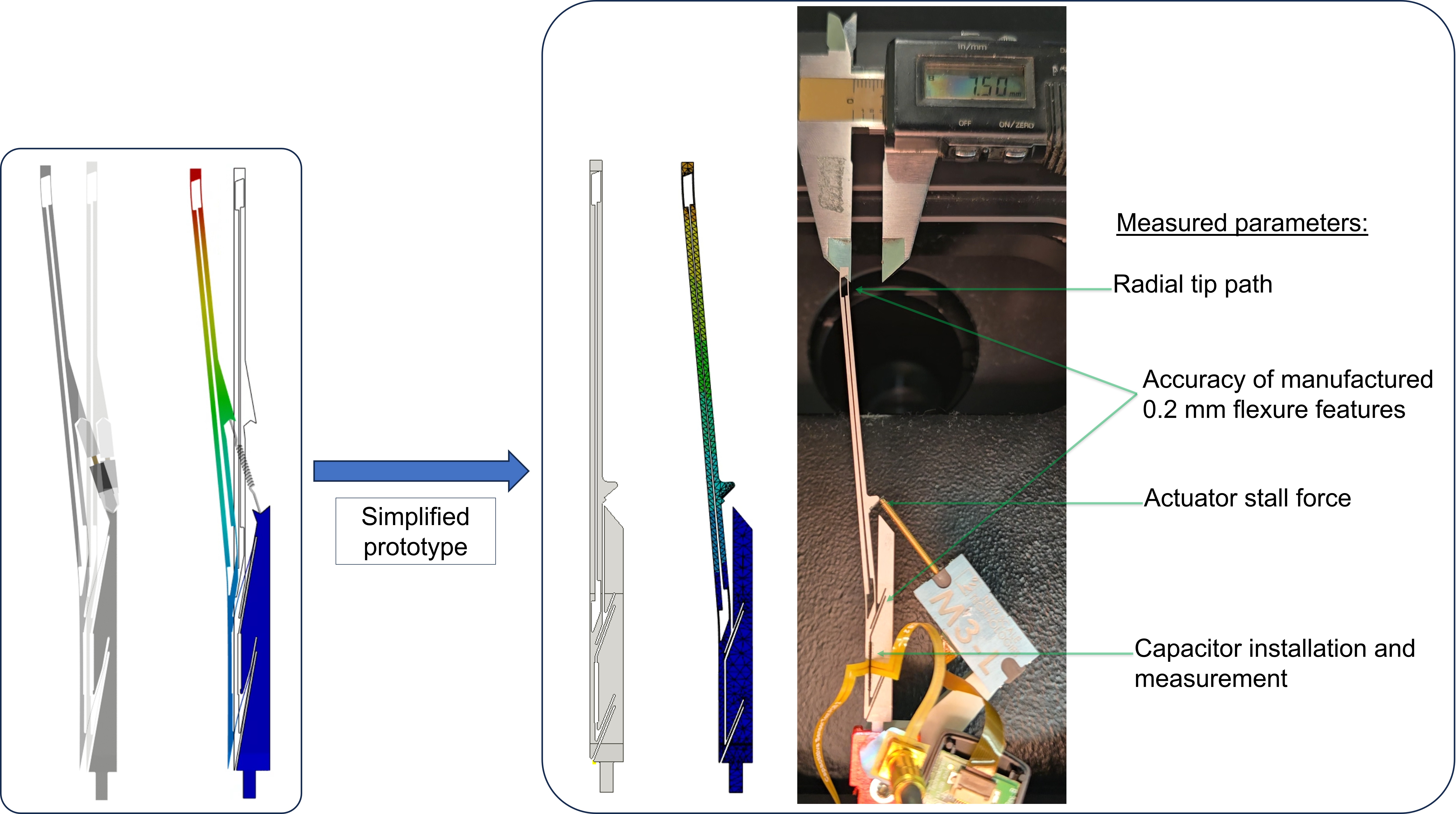}
    \caption{Manufactured flexure with simplified geometry allowing for testing the manufacturing feasibility, instalment/testing of capacitors.}
        \label{fig:ukatc-Labtest}
\end{figure}

A prototype flexure was manufactured using the wire-EDM process from 6082 T6 aluminium alloy. This was designed to enable testing with a development kit version of the radial actuator, so does not include all the interface geometry described above, nor does it reflect the current design and so has a significantly higher defocus error than $50\,\mu\mathrm{m}$. However, it was useful in demonstrating that the flexure geometry is manufacturable, that the path closely follows theoretical path with good focus compensation and also verified the required actuation force and possibility to integrate capacitor sensors as shown in Figure ~\ref{fig:ukatc-Labtest}.

\FloatBarrier

To support quick design iteration, FEA simulations using Ansys were used. The validation of the FEA model was performed experimentally on an aluminium flexure with double parallelogram as seen in Figure 7. The resulting difference between the FEA and experimental model result in terms of defocus was 5\%.

\begin{figure}[ht]
    \centering
    \includegraphics[width=0.75\linewidth]{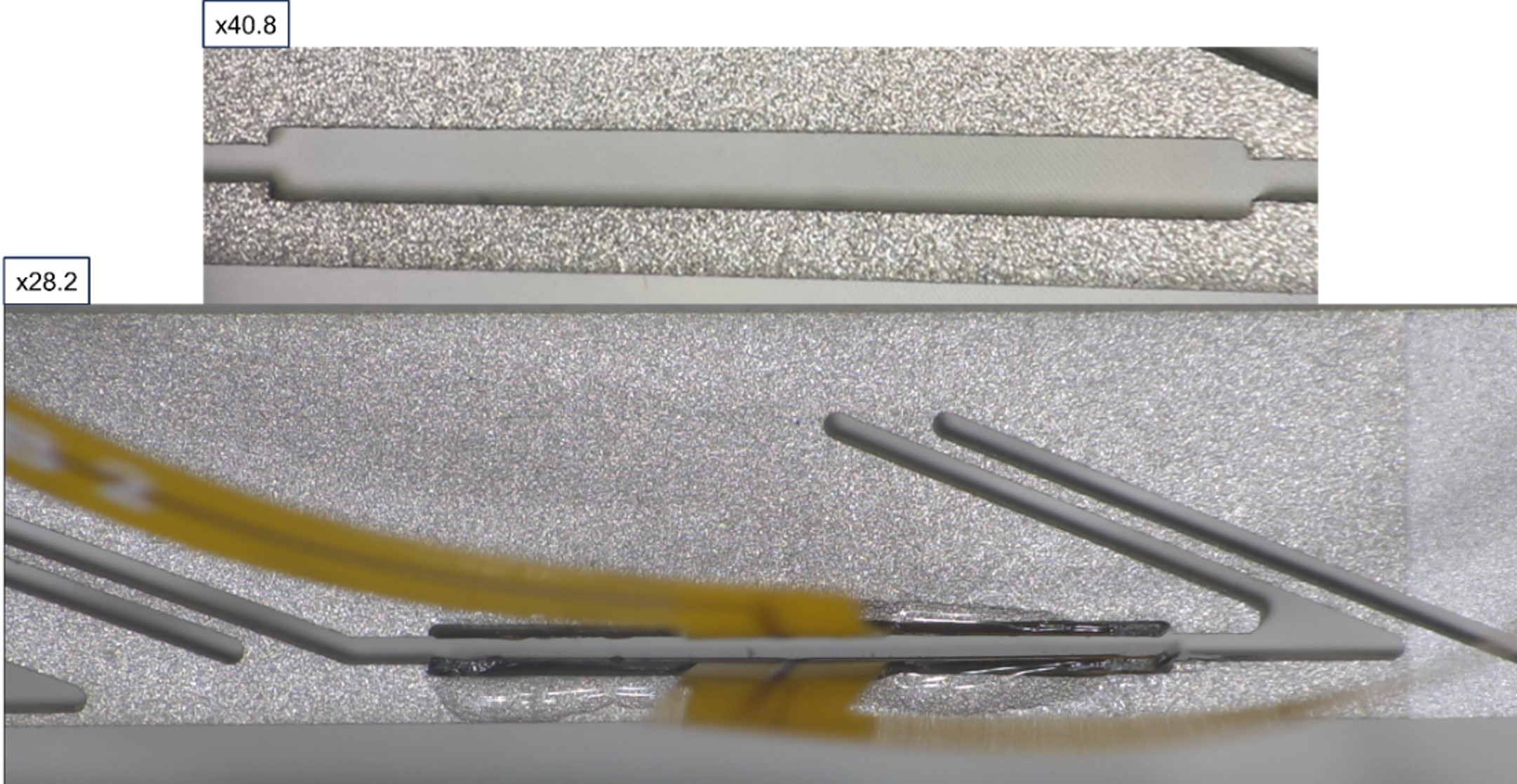}
    \caption{Flex capacitor PCB (right), capacitor pockets before assembly (upper) and after assembly (lower). The capacitor sensors are 2.6mm x 8.8mm hosted in a board that is 3.6mm x 9.8mm.}
        \label{fig:ukatc-capacitors}
\end{figure}

The radial sensor comprises two active electrodes mounted within the compensator flexure, seen in Figure ~\ref{fig:ukatc-capacitors}, measuring electrode separation displacements of approximately 1~mm. Prior to actuation, the gap distance is $300\,\mu\mathrm{m}$ with a calculated capacitance of 0.68 pF. Following full actuation, the gap increases to $1300\,\mu\mathrm{m}$ with a corresponding capacitance of 0.16~pF, yielding a non-linear capacitance range of approximately 0.16 to 0.68 pF. A measurement sensitivity of 50~aF is required for the radial sensor, which, while demanding, remains within the capability of the selected sensing chip. If greater sensitivity is needed, this could be achieved by either reducing the initial electrode separation or increasing the electrode area

\FloatBarrier

\subsection{Tilting spine Positioner concept}
\begin{figure}[ht]
    \centering
    \includegraphics[width=\linewidth]{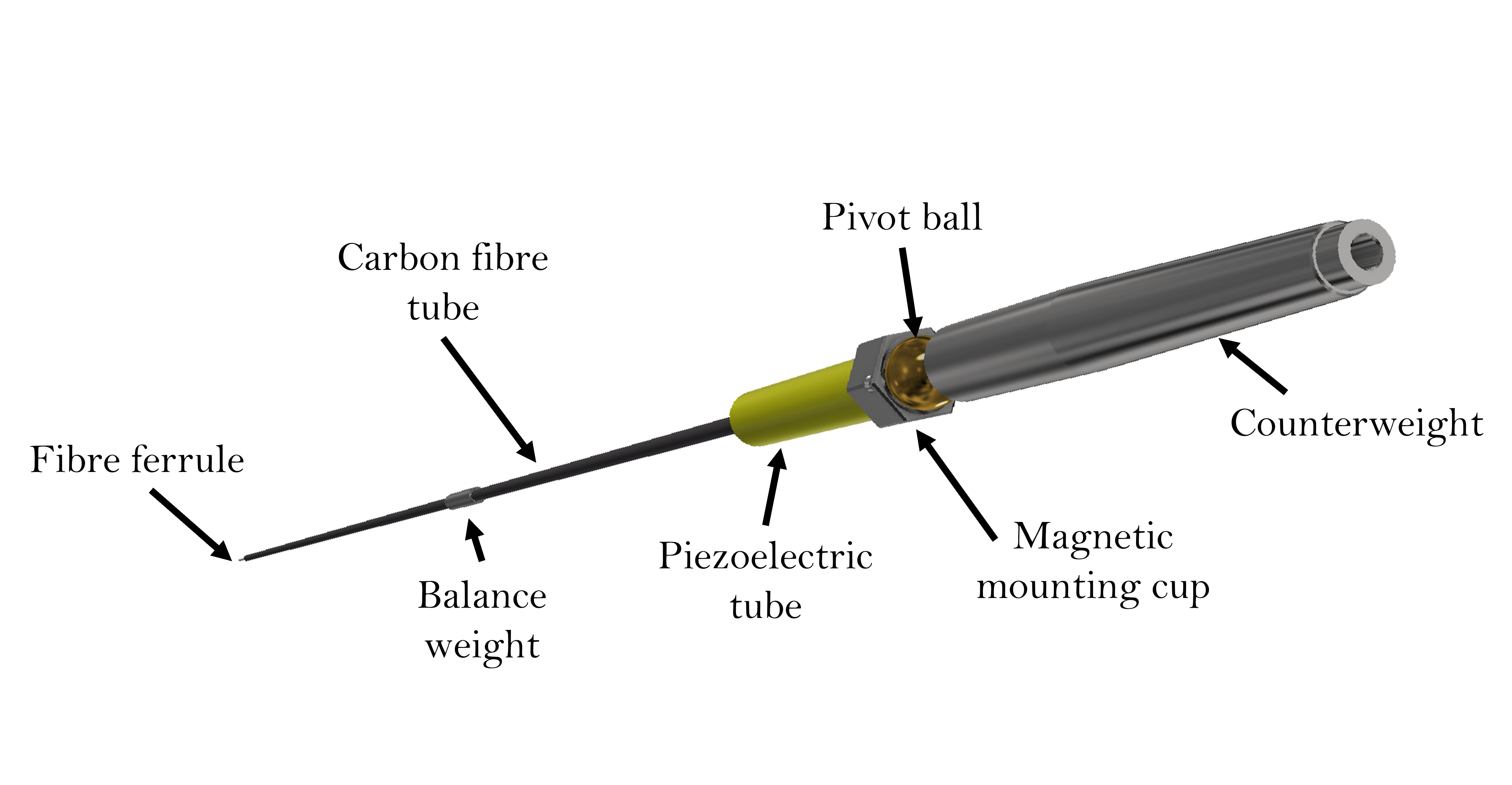}
    \caption{Labeled tilting spine fibre positioner}
    \label{fig:AAO_Tilting_spine}
\end{figure}

The tilting spine positioner concept for WST developed by Australian Astronomical
Optics (AAO) is the evolution of an established fibre positioner concept first developed
for the Subaru Telescope (FMOS-Echidna) and more recently for the VISTA instrument
4MOST (AESOP). This robust, low part count fibre positioner functions through the
stick-slip rotation of a pivot ball by a piezoelectric actuator, which in turn moves a
carbon fibre ferrule housing the optical fibre, and can be seen in Figure \ref{fig:AAO_Tilting_spine}. Due to the fibre routing through a
central carbon fibre ferrule, this design offers a positioner that is both robust to
collisions whilst also reducing additional stresses on the fibre. Through using small
diameter ferrules, 0.5 mm, the tilting spine positioner offers minimal dead zones
within focal plane and the ability for high clustering of fibres due to a 0.7 mm closest
approach between fibres and a moderate patrol radius of 1.4x pitch, with a higher
patrol radius design being developed. In order to mitigate the tilt and defocus losses
that can occur at large patrol radii of a tilting spine, a 1.9° limit is placed on the tilting spine and an increased spine length is chosen.
\FloatBarrier

As part of the evolution from previous tilting spine positioners, the advances in the
positioner focus on the reduction of the pitch of the fibre positioner along with the
footprint whilst aiming to increase the patrol radius towards a 2x pitch, which would
allow for a 1:16 packing structure of High Resolution (HR) fibres relative to Low
Resolution (LR) fibres whilst maintain full coverage of the focal surface. Due to the
vast multiplexing requirements a reduction of the operating voltage and power
requirements during repositioning is key focus of the tilting spine evolution.

\subsection{Theta-Phi Positioner Concept}
\label{subsec:theta-phi}

Drawing on the Astrobots laboratory's experience with the SDSS-V, DESI, and MOONS
projects, we adopt a well-proven two-axis rotational architecture, commonly referred to
as a SCARA\footnote{Selective Compliance Assembly Robot Arm.} robot.

Each fibre positioner consists of two rigid arms mounted in series at a fixed base point on the
focal plate, with exactly two rotational degrees of freedom operating in the focal plane:
rotation of the inner arm (length $l_\alpha = 1.8$~mm) about the fixed base axis
perpendicular to the focal surface ($\theta$ axis), and rotation of the outer arm
(length $l_\beta = 1.8$~mm) relative to the inner arm about the elbow joint ($\phi$
axis). The optical fibre ferrule is attached at the outer tip.

\begin{figure}[ht]
    \centering
    \includegraphics[width=\linewidth]{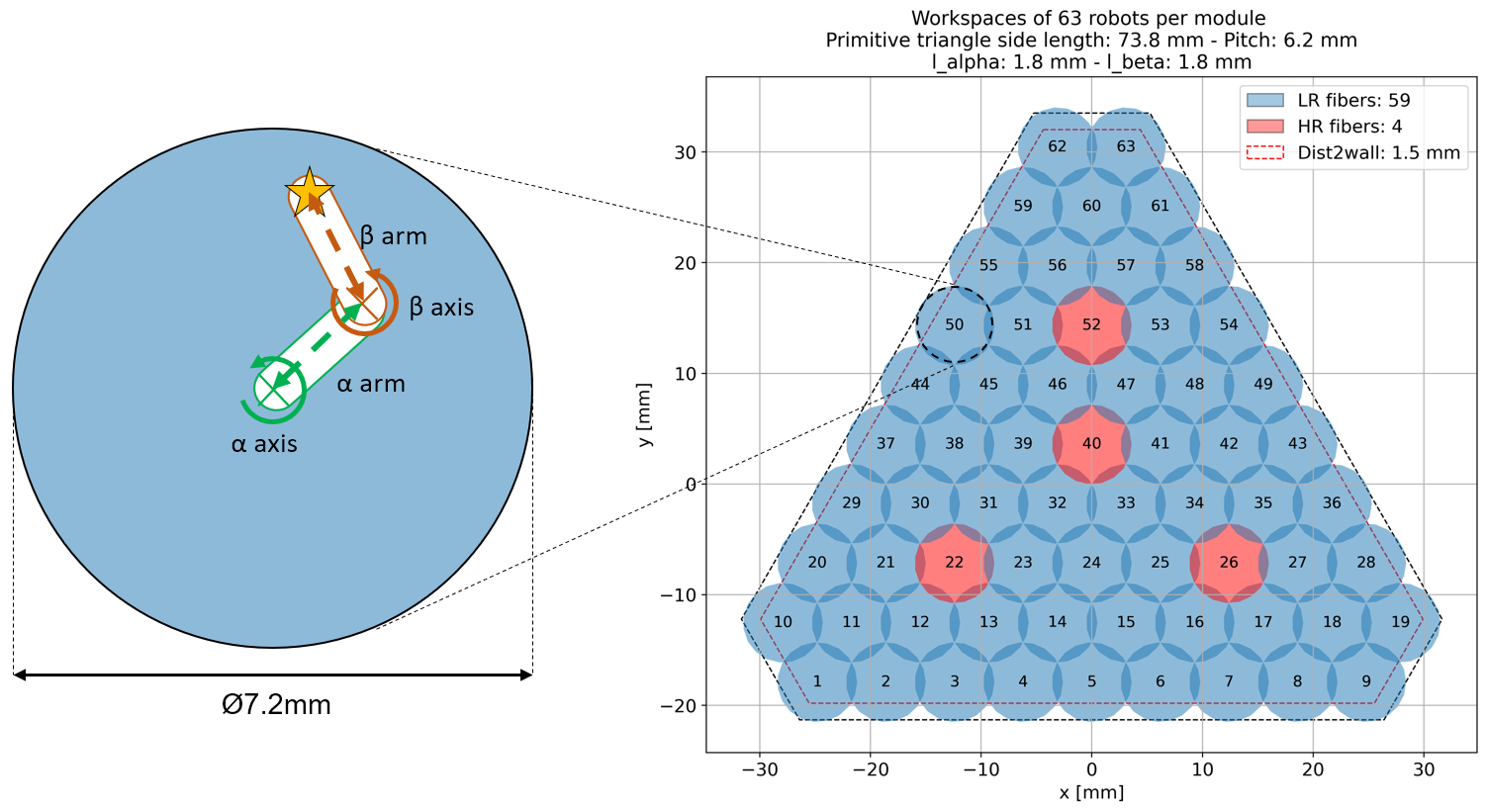}
    \caption{Positioner workspace for the Theta-Phi design}
    \label{fig:pos_workspace}
\end{figure}

With $l_\alpha \approx l_\beta \approx p/2$ at the WST pitch $p = 6.2$~mm, the fibre
tip can reach any point within a 7.2~mm diameter patrol disk centred on its base.
Adjacent patrol zones overlap by approximately 1~mm, ensuring continuous sky coverage
across the focal plane.  The minimum target separation between
neighbouring positioners is currently 2.2~mm (11.3~arcsec at the WST plate scale of
0.195~mm/arcsec), with an ongoing effort to reduce this to 1.5~mm (7.6~arcsec) through
optimised beta-arm geometry and 0.5~mm outer-diameter ferrules as represented on ~\ref{fig:pos_workspace}.

\FloatBarrier

Each positioner is driven by two 4~mm brushless DC motors with planetary gearheads (gear ratios 280:1 to 337:1, depending on motor supplier). The motors are driven open-loop via sensorless field-oriented control, with no Hall sensors or shaft encoders; datum calibration is performed by detecting hard-stop contact through DC-link current monitoring. The control electronics use a shared-board architecture in which a single PCB drives 21~positioners simultaneously via dedicated microcontrollers, one per positioner, communicating over a CAN~2.0 bus at 1~Mbit/s. Three such boards populate
each 63-positioner module.

\subsubsection*{Dual-manufacturer strategy}

\begin{figure}[ht]
    \centering
    \begin{subfigure}[b]{0.48\textwidth}
        \centering
        \includegraphics[width=\textwidth]{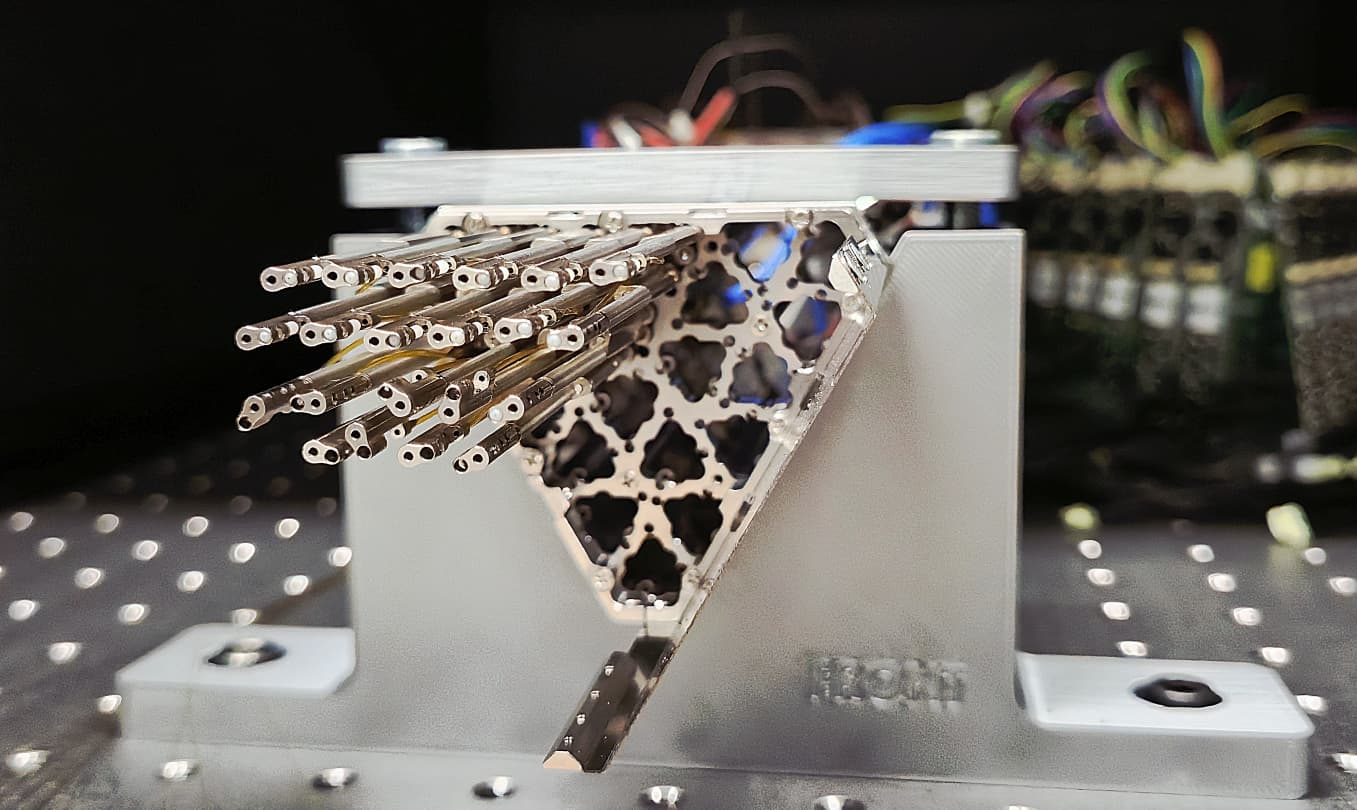}
        \caption{21 positioners Trillium design, manufactured by Orbray LTD.}
        \label{fig:Orbray21}
    \end{subfigure}
    \hfill 
    \begin{subfigure}[b]{0.48\textwidth}
        \centering
        \includegraphics[width=\textwidth]{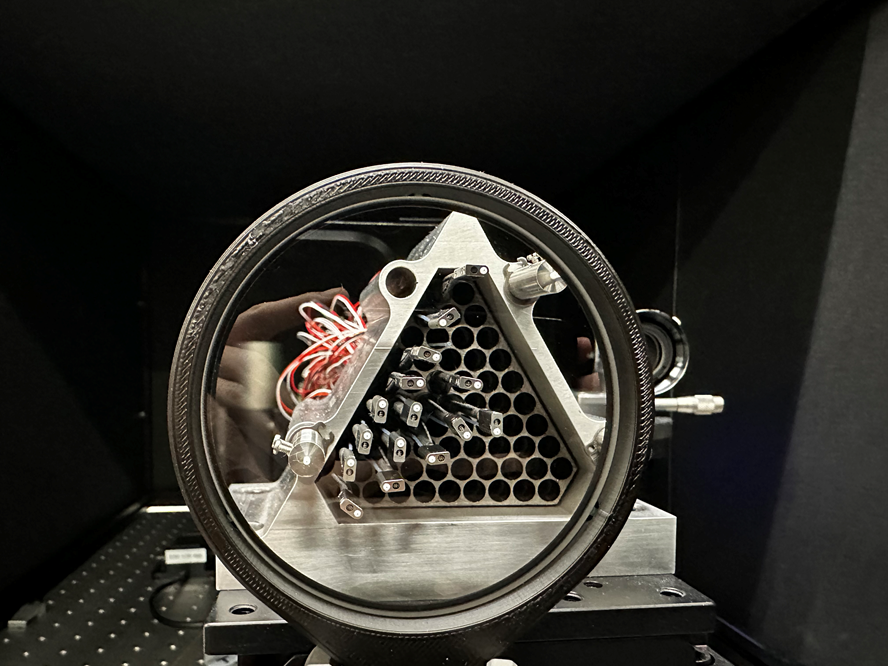}
        \caption{21 positioners, manufactured by MPS AG}
        \label{fig:MPS21}
    \end{subfigure}
    \vspace{10pt}
    \caption{The two design in competition with different suppliers (a) Orbray LTD manufacturer (b) MPS AG manufacturer.}
    \label{fig:DualSupply}
\end{figure}

To mitigate single-supplier risk at the 32\,000-unit production scale, two mechanical implementations are being developed in parallel. The first, manufactured by MPS Micro Precision Systems (Switzerland) as the 21 protype represent on Figure \ref{fig:MPS21}, uses individually assembled robots with independent alpha and beta axes mounted in a shared triangular chassis. The second, manufactured by Orbray Co., Ltd.\ (Japan) as the 21 protype represent on Figure \ref{fig:Orbray21}, follows the open-source Trillium design from LBNL,\cite{Trillium2024} in which positioners are grouped in sets of three with mechanically coupled axes requiring software compensation when the alpha arm rotates.

\subsubsection*{Prototype testing}

Both manufacturers have delivered functional prototypes that have undergone systematic
 characterization in EPFL.\cite{Galal2025,PinedaSuarez2025} A six-positioner MPS
prototype achieved alpha-arm repeatability below
$1.2\,\mu\mathrm{m}$~RMS, with datum repeatability below
$10\,\mu\mathrm{m}$ and $2\,\mu\mathrm{m}$, respectively, well within the WST
specifications. Lifetime testing on this prototype exceeded 300\,000 moves per arm with
minor degradation in repeatability overtime. Thermal testing from $-20\,^\circ$C to $+30\,^\circ$C confirmed nominal
performance across most of the range, with two units showing motor anomalies at the
coldest temperatures currently under investigation. 

A 21-positioner Orbray prototype based on the Trillium design was subsequently tested.
Beta-arm repeatability was mostly within the $6\,\mu\mathrm{m}$ specification, while
alpha-arm repeatability showed greater variability, with non-smooth motion in several
units attributed to mechanical coupling effects within the trillium assembly. Beta-arm
backlash ($7^\circ$--$17^\circ$) exceeded the $5^\circ$ specification and is being
addressed through software calibration and mechanical refinement. Angular tilt averaged
$0.49^\circ$ against a $0.4^\circ$ target. Overall, the Orbray prototype represents an
improvement over the previous iteration but requires further work, particularly on
alpha-arm motion quality and beta backlash. A direct comparison with a 21-positioner MPS
prototype is planned for the next testing campaign.
 
\subsubsection*{Module Organisation}
 
Individual positioners are grouped into modules, each driven by a dedicated local
controller that handles motor commutation, path planning, collision avoidance within
the module, and fault detection, see Figure \ref{fig:elec_architecture}. As a reference, the EPFL triangular-module concept
groups 63 positioners per module; 528 such modules tile the curved focal surface,
yielding a total multiplex of 33\,264. Modules communicate with the central instrument
control system for target distribution, metrology data exchange, and health monitoring.
This modular architecture facilitates parallel integration and testing, and allows
individual units to be replaced without disturbing the remainder of the focal plane.

\section{Modularity concept}
\label{sec:ModularityConcept}

Given the large number of positioners required to populate the focal plane, innovative modular solutions are necessary to address the key challenges of assembly, integration, and long-term maintenance. Modularity is therefore not merely a design convenience but a driving requirement: it must simplify the assembly process, enable efficient module-level testing prior to integration, and provide a practical strategy for in-service replacement or repair of individual modules without disturbing the surrounding focal plane.

Two main modular approaches emerged from the team design studies. The first is an  \textbf{inline arrangement}, in which positioners are grouped in linear modules along a common axis. The second is a \textbf{triangular arrangement}, in which positioners are grouped following a triangular lattice pattern. Both approaches offer distinct trade-offs in terms of focal plane coverage, packing efficiency, mechanical complexity, and serviceability, which are assessed in the evaluation process.

\subsection{Inline (curvilinear) module}
The inline modules were designed using a modified triskele arrangement to generate curvilinear segments that tessellate together to form a hexagonal focal surface outline with uniform spherical curvature. This geometry also produces a central hexagonal cutout to accommodate the Integral Field Spectrograph (IFS) pick-off mirror. 

Only two identical module types are required to populate the entire curved focal surface, allowing for streamlined manufacturing and reduced assembly complexity for the 30,000+ positioner system.

\begin{figure}
    \centering
    \includegraphics[width=\linewidth]{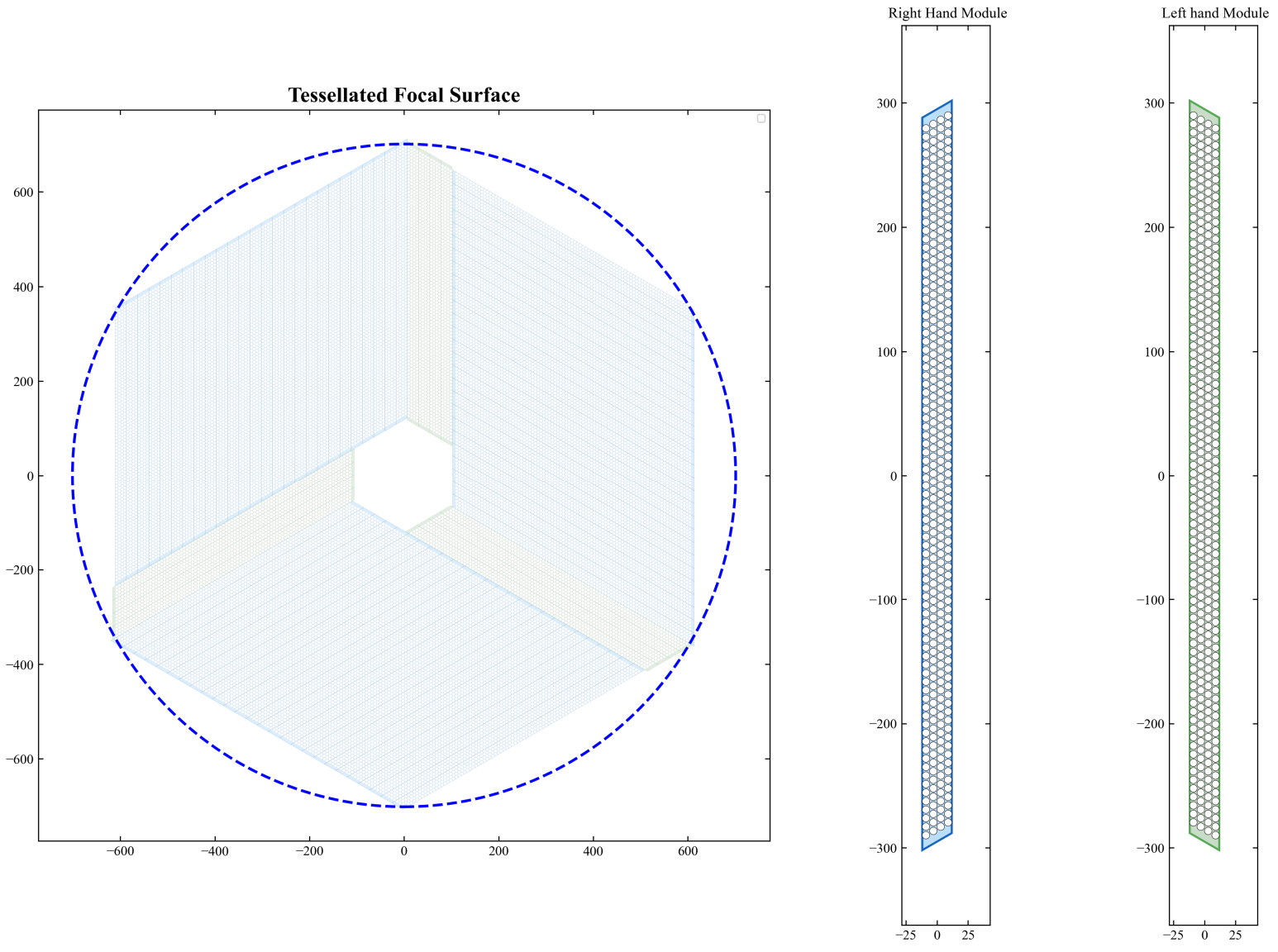}
    \caption{Fully tiled focal surface layout for the in-line curvilinear module design (left), and respective right- and left- handed modules (right)}
    \label{fig:inline_FP}
\end{figure}

\begin{figure}
    \centering
    \includegraphics[width=\linewidth]{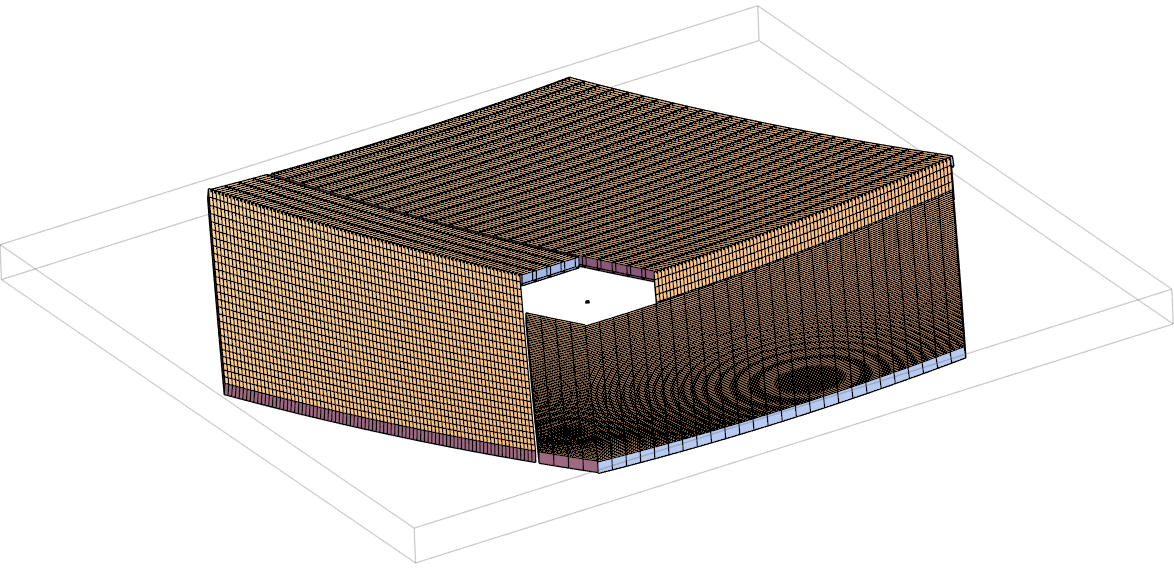}
    \caption{3D view of the Inline focale plane concept}
    \label{fig:inline_FP_2}
\end{figure}

The module design also facilitates mounting alongside structural y-struts integrated under the focal plate, providing both structural rigidity to the system and ensuring that pitch spacing is preserved across module boundaries, effectively minimizing dead space across the full focal surface as represented on Figure ~\ref{fig:inline_FP} and on Figure ~\ref{fig:inline_FP_2}.

\FloatBarrier
\FloatBarrier

\FloatBarrier
\FloatBarrier

Each module utilizes a 84x4 fiber layout for a total of 336 fibers. This layout was chosen to both accommodate a 1:16 HR fiber packing structure to ensure full HR coverage of the focal surface as well as to improve modularity with fiber designation to spectrographs.

\subsection{Triangular module}
\label{subsec:triangular-module}

\begin{figure}[ht]
    \centering
    \includegraphics[width=\linewidth]{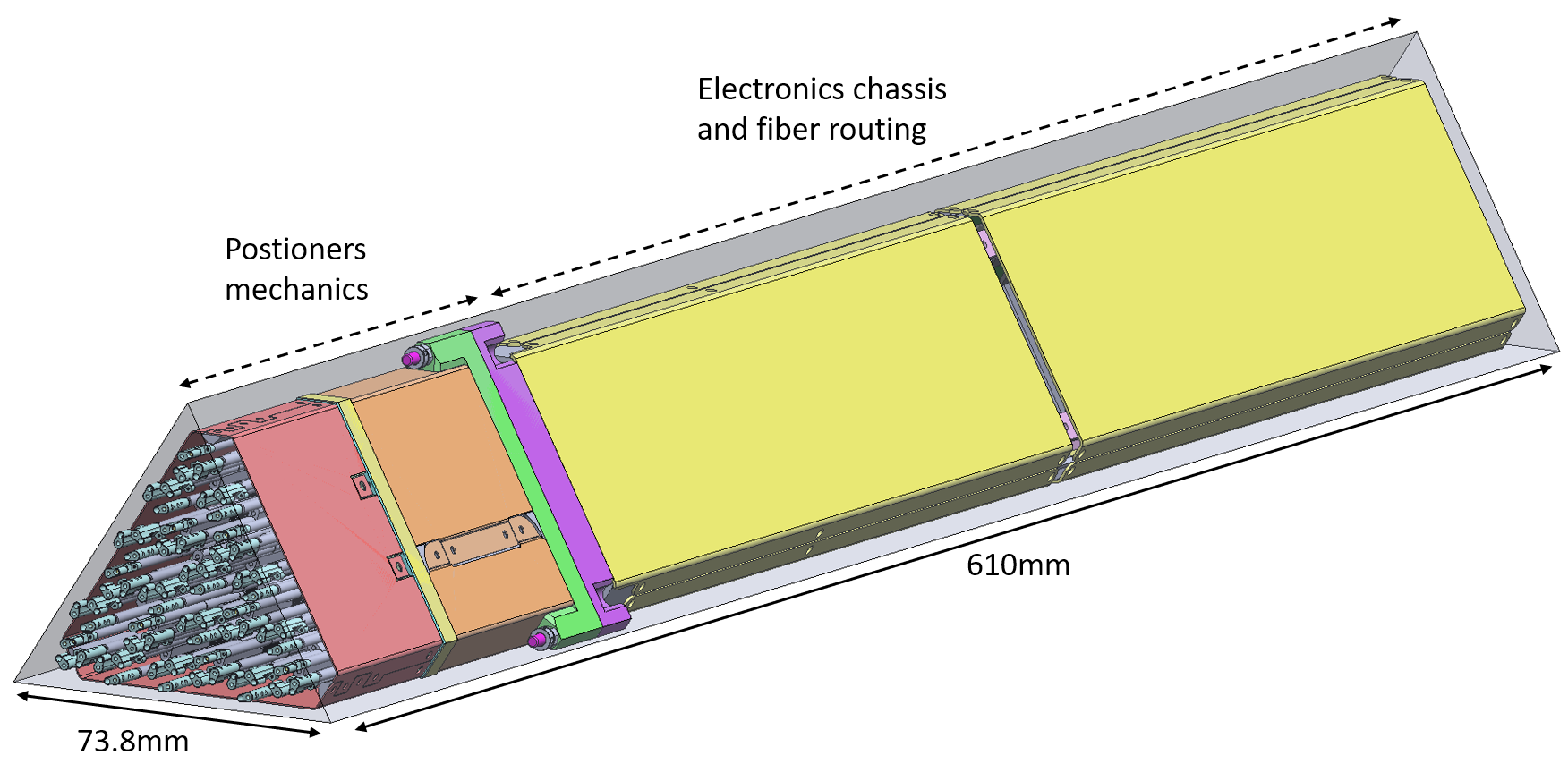}
    \caption{Module with electronic PCB placement}
    \label{fig:overview_module_envelope}
\end{figure}

The triangular module concept, developed jointly by EPFL and UKATC, groups 63
positioners into a compact equilateral-triangle chassis. EPFL used a  side length of 73.8~mm as UKATC has a slightly bigger triangle footprint, and an overall module length of approximately 610~mm (see Figure  ~\ref{fig:overview_module_envelope}\cite{Rombach2024}). The 63-unit
count was optimised to balance three competing constraints: sufficient positioner density for high focal-surface coverage, compatibility with the curved focal surface geometry, and practical limits on module-level assembly and testing complexity.\cite{Galal2025}

\FloatBarrier

\begin{figure}[ht]
    \centering
    \includegraphics[width=\linewidth]{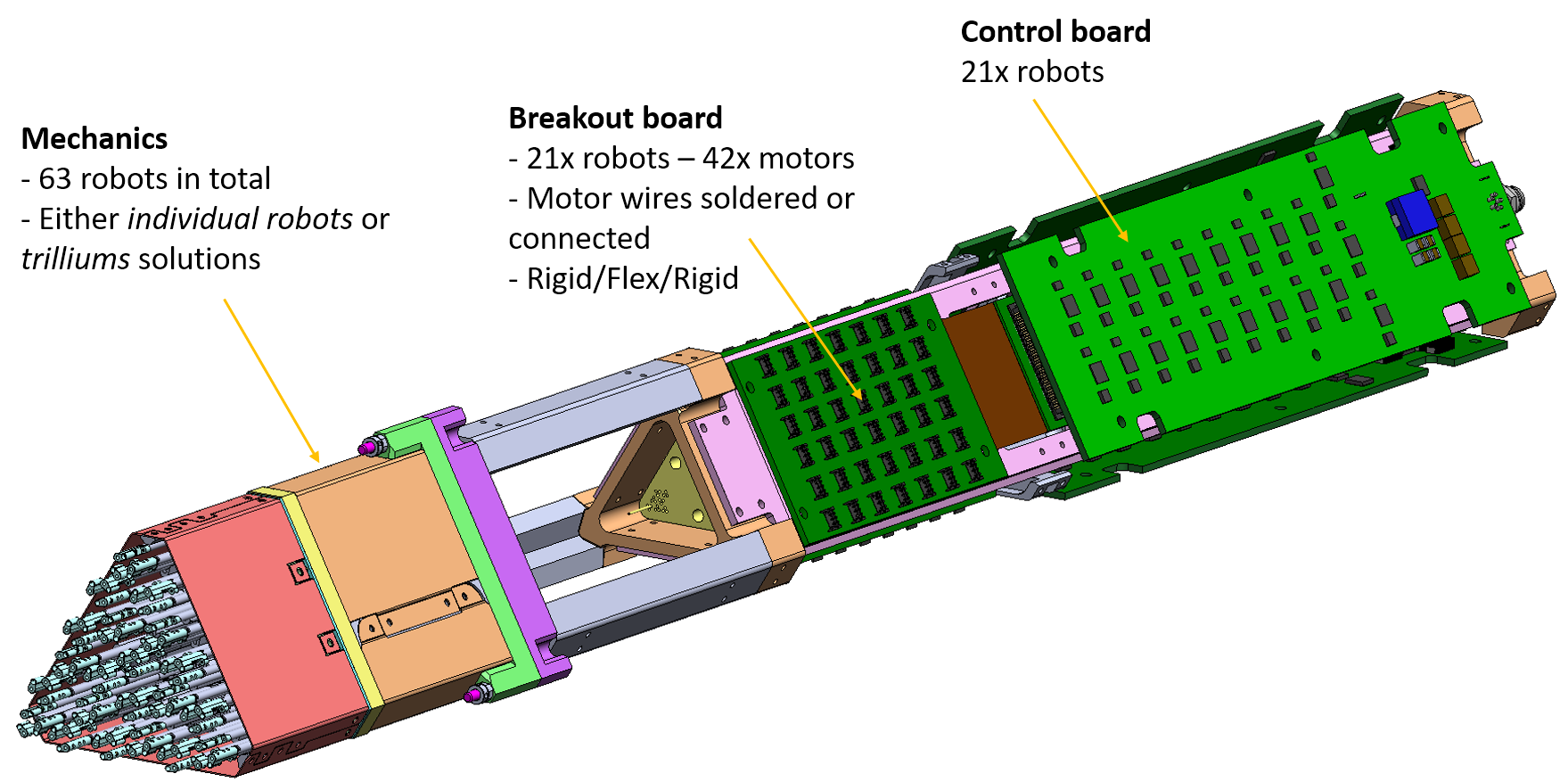}
    \caption{Module with electronic PCB placement}
    \label{fig:elec_architecture}
\end{figure}

The positioners are arranged on a triangular lattice at the nominal 6.2~mm pitch. Each
module is internally subdivided into three groups of 21~positioners, each driven by a
dedicated shared control PCB~\cite{Pernecker2026electronics} communicating over a CAN~2.0 bus (see
Section~\ref{subsec:theta-phi}) . This $3 \times 21$ architecture enables parallel
module-level integration: all 63~positioners can be functionally tested prior to
installation in the focal plane. Four fibre positions per module are reserved for
high-resolution (HR) fibres in a 1:16 packing ratio relative to low-resolution (LR)
fibres, ensuring full HR coverage of the focal surface as represented on Figure ~\ref{fig:elec_architecture}.

\FloatBarrier


\subsubsection*{Focal plane tiling}

Approximately 520 triangular modules tile the 1\,500~mm diameter curved focal surface (field curvature radius $R = 6\,021$~mm), with an optimised ``enhanced hexagonal'' layout comprising 516~modules for a total of 32\,508~positioners and a low-resolution sky coverage of 70.1\% of the vignetting disk. 

\begin{figure}[ht]
    \centering
    \includegraphics[width=\linewidth]{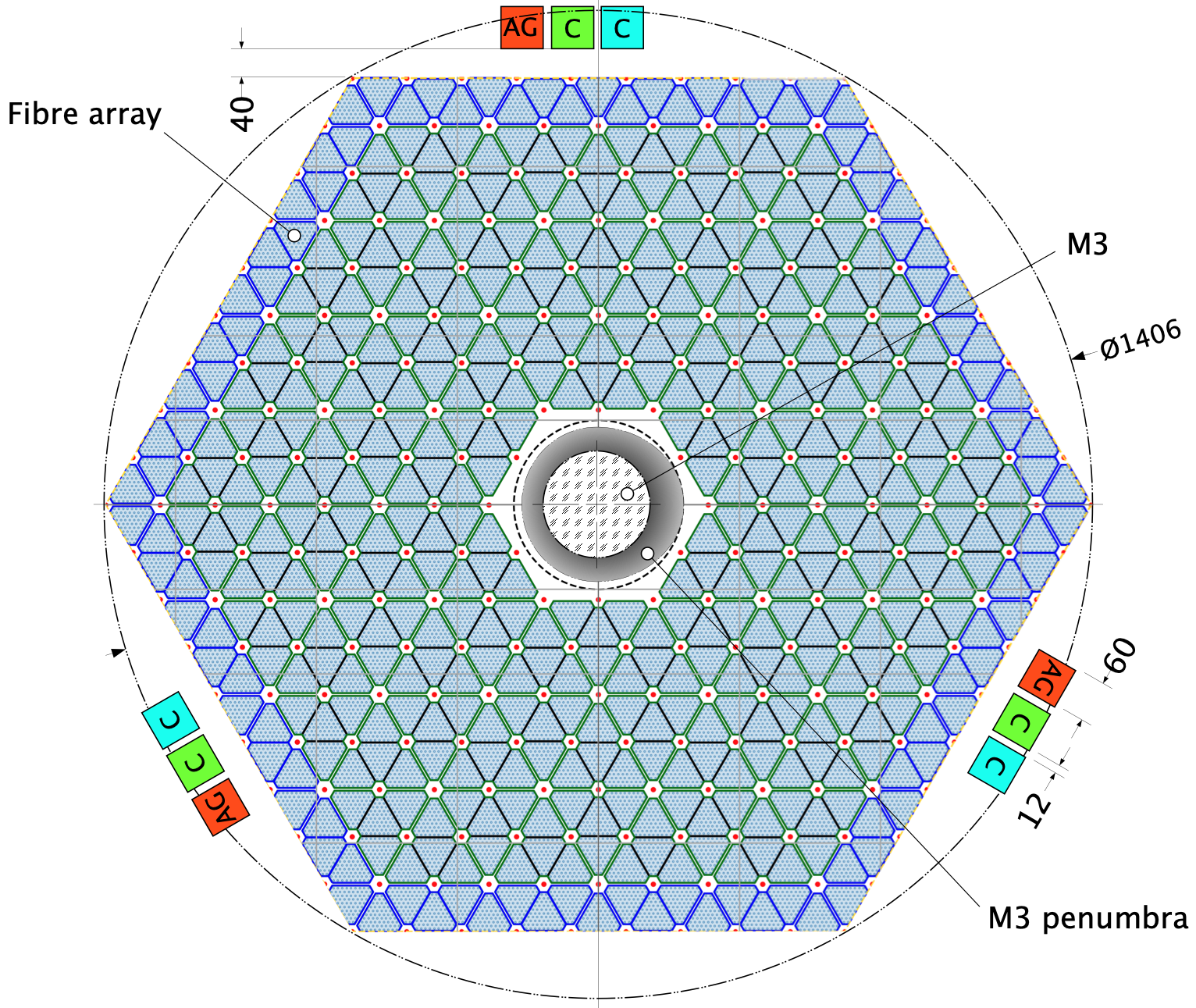}
    \caption{Module layout on the focal plane for the EPFL-UKATC concept}
    \label{fig:layout_FP}
\end{figure}

The layout includes a central exclusion zone (up to 222~mm diameter) to accommodate the Integral Field Spectrograph (IFS)
pick-off mirror and the M3 obstruction as represented on Figure ~\ref{fig:layout_FP}.

The triangular geometry allows
straightforward tessellation of the curved surface with minimal dead space at module
boundaries, while preserving the 6.2~mm pitch across adjacent modules.

\FloatBarrier


\subsubsection*{Focal plate and module mounting}

\begin{figure}[ht]
    \centering
    \includegraphics[width=\linewidth]{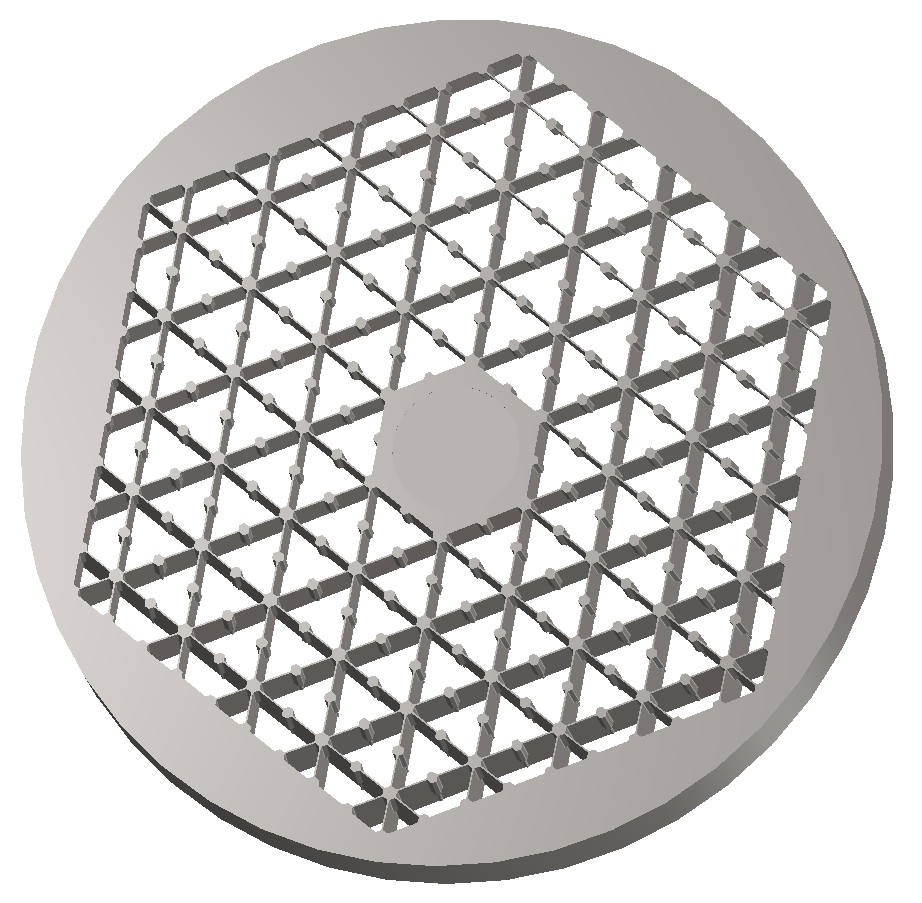}
    \caption{Semi frameless focal plate}
    \label{fig:semi-frameless}
\end{figure}

The modules are mounted onto a monolithic focal plate machined from aluminium 7022
(alternative: stainless steel), with dimensions of 1\,500~mm diameter and 120~mm
thickness ~\ref{fig:semi-frameless}. Finite-element analysis predicts a maximum static deformation at zenith of approximately 35~$\mu$m for aluminium (15~$\mu$m for steel), with a static tilt
contribution below 16~arcsec, negligible within the overall error budget. A 1/6-scale
aluminium prototype has been manufactured and measured on a coordinate measuring
machine, yielding $\pm30\,\mu$m defocus and $\pm0.05^\circ$ tilt, confirming that
the required manufacturing precision is achievable.

\begin{figure}[ht]
    \centering
    \includegraphics[width=\linewidth]{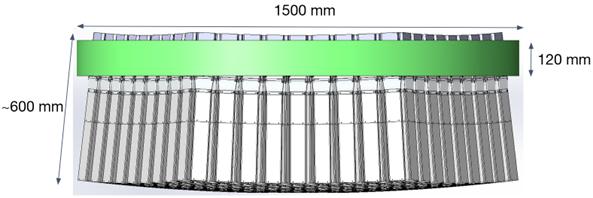}
    \caption{Focal plane side view with populated modules}
    \label{fig:FP_Layout_side_view}
\end{figure}

\begin{figure}[ht]
    \centering
    \includegraphics[width=\linewidth]{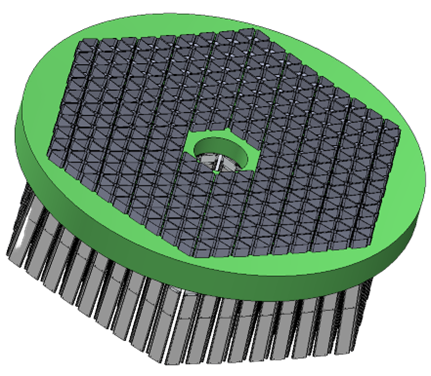}
    \caption{Focal plane with populated modules}
    \label{fig:FP_Layout_side}
\end{figure}

Each module is attached to the focal plate by three M2.5 captive screws with a custom
shim stack that allows adjustment of both tilt and defocus during integration. At the
module edge, the residual defocus with respect to the curved focal surface reaches
approximately 78~$\mu$m and the maximum tilt is $0.29^\circ$; these values are being
mitigated by either fabricating fibres at different focal heights within the module or
shifting the module reference plane by approximately 25~$\mu$m upward  as represented on Figure ~\ref{fig:FP_Layout_side} and on Figure  ~\ref{fig:FP_Layout_side_view}.

\FloatBarrier

\FloatBarrier

\FloatBarrier

\subsubsection*{Maintainability}

The modular architecture is designed for rapid in-service replacement. A single module
can be removed by disconnecting its electrical and fibre cables and unscrewing the three
mounting fasteners. Individual positioner failures are handled by the control software,
which defines an avoidance zone around the faulty unit without requiring physical module
replacement. Full module swap-out is triggered only in the event of a control-board
failure or a large number of degraded positioners within the same module.

\section{Trade off study}
\label{sec:TradeOffStudy}
Since we have several designs, we need to evaluate them to present two designs at ESO. The first design selected will be with low TRL but high coverage, the second one will have a higher TRL but with lower coverage than the first selected design. 

\begin{figure}[ht]
    \centering
    \includegraphics[width=1\linewidth]{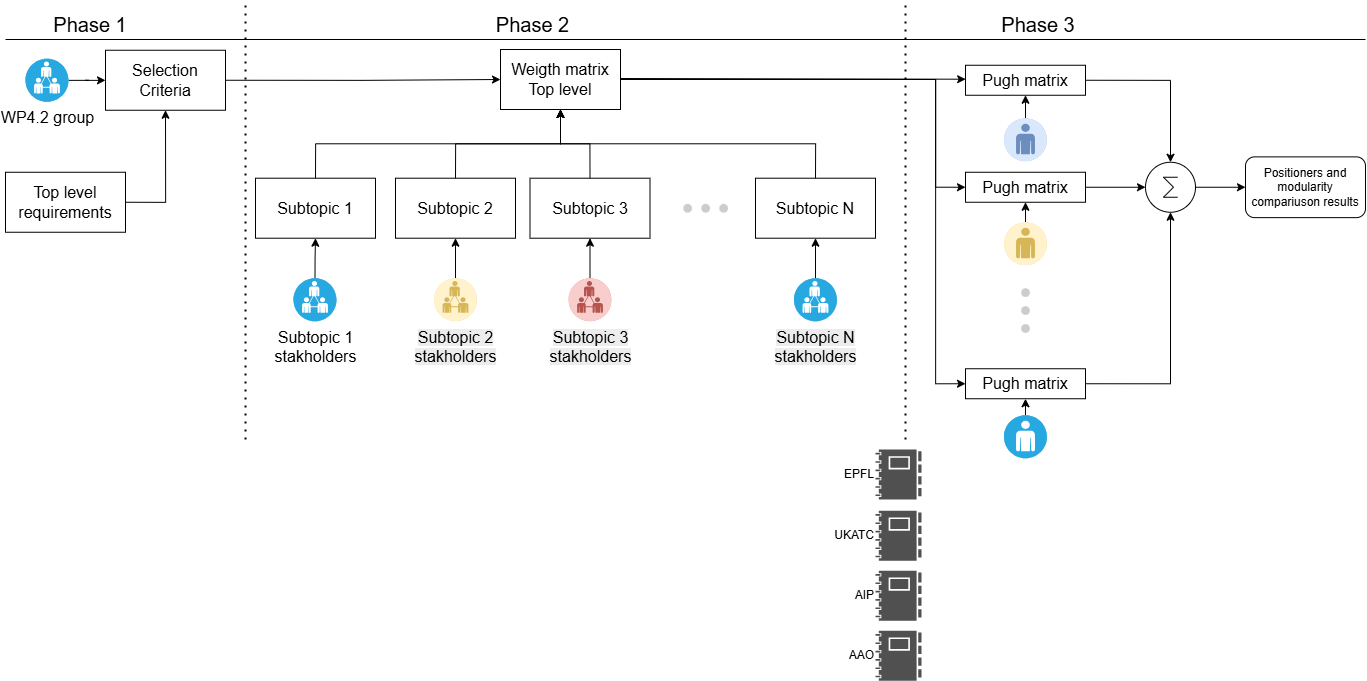}
    \caption{Workflows of the selection process}
    \label{fig:process_overview}
\end{figure}

We define a three-phase process to ensure a proper evaluation that minimizes the impact of individual reviewer biases. Figure~\ref{fig:process_overview} represents the full process of weighting and the selection of the best design over all team proposals.

\subsection{Evaluation criteria selection}
To help the WST consortium select the appropriate design, the WP4.2 positioner team first defines a list of selection criteria to evaluate the positioner designs. This list takes input from top-level project documents already released, as well as inputs from the whole WP4.2 positioners team.

The list of selection criteria is then grouped into different categories, each identified by a single top-level title. This grouping is required to comply with the data format of the subsequent weighting process, and serves to assign dedicated stakeholder groups responsible for evaluating the weight of each category of criteria, as described in Section~\ref{sec:weighting}.

\FloatBarrier

Once the WP4.2 positioners team agrees on the criteria and their grouping, the first phase is validated and the process moves on to Phase 2: the weight evaluation.

\subsection{Evaluation criteria weighting}
\label{sec:weighting}
This process is split across the different categories of selection criteria, allowing a group of reviewers representative of each specific category to be selected independently. A reviewer may belong to more than one stakeholder group, and this granularity allows every member of the WST project to be involved or not without impacting the overall process.

The weighting method used is the Analytic Hierarchy Process (AHP)~\cite{Saaty1980}, implemented via a dedicated Excel workbook developed by Klaus D.~Goepel. Each stakeholder receives a link to the Excel sheet and fills in one page named after their family name. The evaluation consists of pairwise comparisons of criteria: the reviewer first indicates which criterion (A or B) is more important, and then provides a confidence scale from 1 to 9, where a low value indicates uncertainty and a high value confirms the comparison.

Individual reviewer weights can also be assigned within each stakeholder group, allowing the WST consortium to reduce biases or apply specific expertise-based considerations.

Once all evaluations are completed, the individual results are automatically consolidated into a Summary page, from which the overall weight matrix for all selection criteria is computed.

\subsection{Evaluation process}
\label{sec:evaluation}
Based on the positioner and modularity concept reports produced by each team following a common document template, every member of the WST project team performs a comparative evaluation of all designs against each selection criterion.

Each reviewer receives a link to an individual Excel sheet~\cite{Goepel2018} in which they grade each design on every criterion of selection, filling only the designated cells. The workbook is protected to guide the review and avoid mistakes. The grading scale is defined as follows:

\begin{table}[h]
\centering
\caption{Grading scale for design comparison}
\label{tab:grading}
\begin{tabular}{cl}
\hline
\textbf{Grade} & \textbf{Meaning} \\
\hline
$-2$ & Appears clearly worse than the standard solution \\
$-1$ & Appears slightly worse than the standard solution \\
$0$  & Appears equal to the standard solution \\
$+1$ & Appears slightly better than the standard solution \\
$+2$ & Appears clearly better than the standard solution \\
\hline
\end{tabular}
\end{table}

\begin{figure}[ht]
    \centering
    \includegraphics[width=1\linewidth]{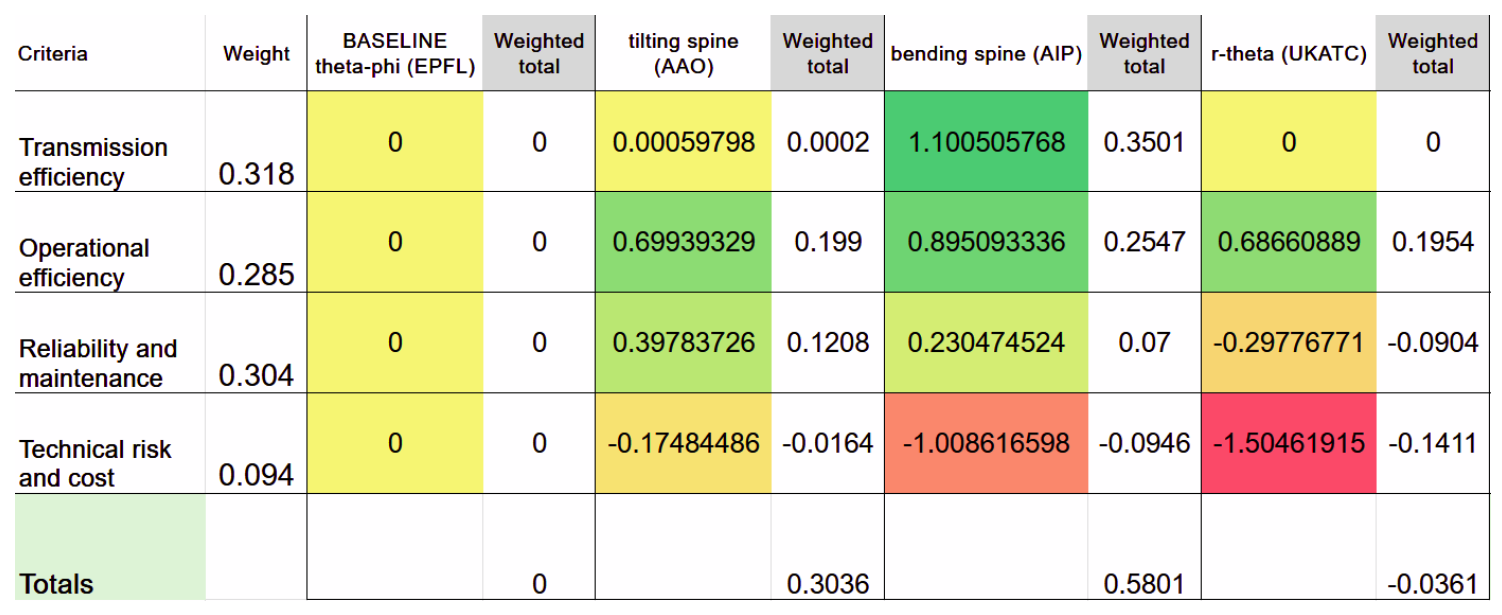}
    \caption{Results of the tradeoff study}
    \label{fig:results_summary}
\end{figure}\FloatBarrier

The weights derived from the AHP process are automatically populated in the comparison sheet, so that each grade is multiplied by the corresponding criterion weight. This allows the weighted total score per design to be automatically calculated and reported on a summary page (see Figure~\ref{fig:results_summary}).

This allow every reviewer to independently record their evaluation, and during the 
consolidation phase, all individual results are aggregated using a Pugh 
matrix~\cite{Pugh1991} approach to produce a final ranking score per positioner design.

\section{CONCLUSION}
\label{sec:Conclusion}

The WST Multi-Object Spectrograph imposes fibre-positioning requirements without precedent: 32,000 robotic units delivering 5\,$\mu$m RMS accuracy within a 6.2\,mm pitch across a 1,500\,mm diameter curved focal surface, representing a more than sixfold increase in scale over current instruments such as DESI and 4MOST. To manage the engineering and industrial-scale production risks inherent at this multiplexing level, WP4.2 adopted a deliberate multi-concept development strategy, prototyping four distinct positioner architectures, the theta-phi SCARA, FLEX nitinol-tube, R-theta, and tilting-spine designs, in parallel with two competing modular arrangements, triangular and inline curvilinear. 

The prototype testing reported here demonstrates that meeting the WST specifications is feasible. The theta-phi SCARA prototypes achieved alpha-arm repeatability below 1.2\,$\mu$m RMS with successful lifetime and thermal characterisation, while the FLEX, R-theta, and tilting-spine concepts each reached significant maturity in telecentricity control, closed-loop sensing, and high-clustering capability, respectively. In parallel, the modular focal-plane studies confirmed that both the triangular and inline curvilinear architectures can tile the curved surface with minimal dead space while preserving pitch across module boundaries, and a 1/6-scale focal-plate prototype validated the required manufacturing precision.

An initial trade-off study has been carried out to compare the candidate designs. Following a three-phase methodology, the Analytic Hierarchy Process (AHP) was used to weight the selection criteria, and a Pugh matrix was applied to score each concept against a common baseline. This first study provides an early ranking of the architectures and, importantly, exercises and validates the evaluation framework itself, highlighting the criteria and performance metrics that most strongly differentiate the concepts.

This initial assessment is, however, a preliminary step. As the prototypes mature and additional characterisation data, including the planned direct comparison campaigns, become available, the evaluation criteria, weights, and grading inputs will be refined accordingly. A second, consolidated trade-off study will therefore be conducted at the end of the HORIZON Europe-funded conceptual study phase, prior to submission to ESO. This final study will inform the down-selection to two preferred concepts, one offering high focal-surface coverage at lower technology readiness and one of higher readiness with reduced coverage, which will then be advanced toward preliminary design, supporting WST's path toward first light in
the early 2040s.

\section{ACKNOWLEDGMENTS}
This project has received funding from the European Union’s Horizon Europe research and innovation program under grant agreement No. 101183153. Views and opinions expressed are however those of the author(s) only and do not necessarily reflect those of the European Union or the European Research Executive Agency (REA). Neither the European Union nor the REA can be held responsible for them.

\bibliography{report} 
\bibliographystyle{spiebib} 

\end{document}